 \def \VersionLong {}
\documentclass[conference]{IEEEtran}
\IEEEoverridecommandlockouts

\usepackage[utf8]{inputenc}
\usepackage[english]{babel}
\usepackage{float}

\usepackage[ruled,vlined,linesnumbered]{algorithm2e}
	\SetKwInOut{Input}{input}
	\SetKwInOut{Output}{output}

\usepackage{subcaption}
\usepackage{tabularx}
\usepackage{paralist} 

\newenvironment{ienumeration}
	{\ifdefined\VersionLong\begin{enumerate}\else\begin{inparaenum}[\itshape i\upshape)]\fi}
	{\ifdefined\VersionLong\end{enumerate}\else\end{inparaenum}\fi}

\usepackage{amsmath} 
\usepackage{amssymb} 
\usepackage{graphicx}
\graphicspath{{figures/}}

\usepackage[misc,geometry]{ifsym} 

\ifdefined\VersionWithComments
	\usepackage{draftwatermark}
	\SetWatermarkText{draft}
	\SetWatermarkScale{3}
	\SetWatermarkColor[gray]{0.9}
\fi

\usepackage[svgnames,table]{xcolor}
\definecolor{darkblue}{rgb}{0.0,0.0,0.6}
\definecolor{darkgreen}{rgb}{0, 0.5, 0}
\definecolor{darkpurple}{rgb}{0.7, 0, 0.7}
\definecolor{darkblue}{rgb}{0, 0, 0.7}

\usepackage[
		colorlinks=true,
	\ifdefined \VersionWithComments
		pagebackref=true,
	\fi
		citecolor=darkgreen,
		linkcolor=darkblue,
		urlcolor=darkpurple,
	]{hyperref}

\usepackage[capitalise,english,nameinlink]{cleveref} 
\crefname{line}{\text{line}}{\text{lines}} 

\newcommand{\defProblem}[3]
{%
\noindent\fcolorbox{black}{blue!15}{
	\begin{minipage}{.95\columnwidth}
		\textbf{#1 problem:}\\
		\textsc{Input}: #2\\
		\textsc{Problem}: #3
	\end{minipage}
}
	
	\smallskip
	
}

\ifdefined \VersionLong
	\newcommand{\LongVersion}[1]{\ifdefined\VersionWithComments{\color{red!40!black}#1}\else#1\fi}
	\newcommand{\ShortVersion}[1]{\ifdefined\VersionWithComments{\color{black!40}#1}\fi}
\else
	\newcommand{\LongVersion}[1]{\ifdefined\VersionWithComments{\color{black!40}#1}\fi}
	\newcommand{\ShortVersion}[1]{\ifdefined\VersionWithComments{\color{red!40!black}#1}\else#1\fi}
\fi

\usepackage{tikz}
\usetikzlibrary{arrows,automata}
\tikzstyle{every node}=[initial text=,font=\footnotesize]
\tikzstyle{location}=[rectangle, rounded corners, minimum size=12pt, draw=black, fill=blue!10, inner sep=2pt]
\tikzstyle{invariant}=[draw=black, dotted, inner sep=1pt] 
\tikzstyle{urgent}=[fill=yellow, double]
\tikzstyle{bad}=[fill=red!50]

\newcommand{\cellHeader}[1]{\cellcolor{blue!20}\textbf{#1}}
\newcommand{\cellNA}{\cellcolor{gray!50}N/A}

\usepackage{amsthm}
\theoremstyle{plain}

\theoremstyle{definition}
\newtheorem{definition}{Definition}
\newtheorem{example}{Example}

\theoremstyle{remark}
\newtheorem{remark}{Remark}

\usepackage{listings}
\usepackage{color}

\definecolor{mygreen}{rgb}{0,0.6,0}
\definecolor{mygray}{rgb}{0.5,0.5,0.5}
\definecolor{mymauve}{rgb}{0.58,0,0.82}

\lstdefinestyle{ariane}{
	backgroundcolor=\color{white},   
	basicstyle=\footnotesize,        
	breakatwhitespace=false,         
	breaklines=true,                 
	captionpos=b,                    
	commentstyle=\color{mygreen},    
	deletekeywords={...},            
	escapeinside={\%*}{*)},          
	extendedchars=true,              
	frame=single,	                   
	keepspaces=true,                 
	keywordstyle=\color{red!70!black}\bfseries,       
	morekeywords={deadline,end,in,is,maf,ms,offset,out,period,processing,reactivity,thread,wcet,when},            
	numbers=left,                    
	numbersep=5pt,                   
	numberstyle=\tiny\color{mygray}, 
	rulecolor=\color{black},         
	showspaces=false,                
	showstringspaces=false,          
	showtabs=false,                  
	stepnumber=1,                    
	stringstyle=\color{mymauve},     
	tabsize=2,	                   
}

\definecolor{weborange}{RGB}{255,165,0}
\lstdefinestyle{imitator}{
	backgroundcolor=\color{white},   
	basicstyle=\footnotesize,        
	breakatwhitespace=false,         
	breaklines=true,                 
	captionpos=b,                    
	commentstyle=\color{mygreen},    
	deletekeywords={...},            
	emph={True},
    emphstyle={\color{blue}\bfseries},
	escapeinside={\%*}{*)},          
	extendedchars=true,              
	frame=single,	                   
	keepspaces=true,                 
	keywordstyle=\color{red!70!black}\bfseries,       
	language=Caml,                 
	morekeywords={automaton,do,end,goto,invariant,loc,stop,sync,synclabs,urgent,when},            
	numbers=left,                    
	numbersep=5pt,                   
	numberstyle=\tiny\color{mygray}, 
	rulecolor=\color{black},         
	showspaces=false,                
	showstringspaces=false,          
	showtabs=false,                  
	stepnumber=1,                    
	stringstyle=\color{mymauve},     
	tabsize=2,	                   
	classoffset=1, 
	otherkeywords={>,<,-,+,&,:,=},
	morekeywords={>,<,-,+,&,:,=},
	keywordstyle=\color{weborange},
	classoffset=0,
}

\ifdefined\VersionWithComments
	\usepackage[colorinlistoftodos,textsize=footnotesize]{todonotes}
\else
	\usepackage[disable]{todonotes}
\fi

\newcommand{\gennote}[3]{\todo[linecolor=#2,backgroundcolor=#2!25,bordercolor=#2]{#3: #1}}
\newcommand{\ea}[1]{\gennote{#1}{blue}{ÉA}}

\newcommand{\jj}[1]{{\gennote{#1}{purple}{JJ}}}

\ifdefined \VersionWithComments
	\newcommand{\todoinline}[1]{\mbox{}{\color{red}{\textbf{TODO}\ifx#1\\\else:\ \fi #1}}} 
\else
	\newcommand{\todoinline}[1]{}
\fi

\ifdefined \VersionWithComments
	\newcommand{\torewrite}[1]{\mbox{}{\color{gray}{\textbf{To rewrite}\ifx#1\\\else:\ \fi #1}}} 
\else
	\newcommand{\torewrite}[1]{}
\fi

\ifdefined \VersionWithReview
	\usepackage{ulem}
	\newcommand{\reviewAdd}[1]{{\color{purple}#1}}
	\newcommand{\reviewDelete}[1]{{\color{red}\sout{#1}}}
\else
	\newcommand{\reviewAdd}[1]{#1}
	\newcommand{\reviewDelete}[1]{}
\fi

\newcommand{\init}{_0}

\newcommand{\A}{\ensuremath{\mathcal{A}}}

\newcommand{\Actions}{\Sigma}
\newcommand{\action}{\ensuremath{a}}

\newcommand{\Clock}{\mathbb{X}} 
\newcommand{\ClockCard}{\ensuremath{{|\Clock|}}} 
\newcommand{\clock}{x} 
\newcommand{\clockval}{w} 
\newcommand{\ClocksZero}{\vec{0}}
\newcommand{\compOp}{\bowtie}

\newcommand{\edge}{e}
\newcommand{\Edges}{E}

\newcommand{\longueflecheRel}[1]{\stackrel{#1}{\mapsto}}

\newcommand{\flecheRel}{{\rightarrow}}

\newcommand{\grandn}{{\mathbb N}}
\newcommand{\grandq}{{\mathbb Q}}
\newcommand{\grandqplus}{\grandq_{+}} 
\newcommand{\grandr}{\ensuremath{\mathbb R}}
\newcommand{\grandrplus}{\ensuremath{\grandr_{+}}} 

\newcommand{\guard}{g}

\newcommand{\invariant}{\mathcal{I}}

\newcommand{\loc}{\ensuremath{\ell}} 
\newcommand{\locinit}{\loc\init}
\newcommand{\Loc}{L} 

\newcommand{\Param}{\mathbb{P}} 
\newcommand{\param}{p} 
\newcommand{\ParamCard}{\ensuremath{{|\Param|}}} 
\newcommand{\pval}{v} 
\newcommand{\R}{{\mathbb{R}}}
\newcommand{\Rgeqzero}{\R_{\geq 0}}

\newcommand{\sinit}{s\init} 
\newcommand{\stopFunction}{\mathcal{S}}
\newcommand{\States}{S} 

\newcommand{\timelapseStop}[3]{#1^{\nearrow + #2}_{\setminus #3}}

\newcommand{\resets}{R}

\newcommand{\reset}[2]{\ensuremath{[#1]_{#2}}}
\newcommand{\valuate}[2]{\ensuremath{#2(#1)}}

\newcommand{\allActExcept}[1]{\ensuremath{\overline{#1}}}

\newcommand{\cheddar}{Cheddar}
\newcommand{\imitator}{\textsf{IMITATOR}}
\newcommand{\uppaal}{\textsc{Uppaal}}

\newcommand{\styleSched}[1]{\ensuremath{\mathsf{#1}}}

\newcommand{\FPS}{\ensuremath{\styleSched{FPS}}}
\newcommand{\RMS}{\ensuremath{\styleSched{RMS}}}

\definecolor{loccolor1}{rgb}{.9, .95, 1}
\definecolor{loccolor2}{rgb}{.9, .95, 1}
\definecolor{loccolor3}{rgb}{.9, .95, 1}
\definecolor{loccolor4}{rgb}{.9, .95, 1}
\definecolor{loccolor5}{rgb}{1, .5, .5} 

	\definecolor{coloract}{rgb}{0.50, 0.70, 0.30}
	\definecolor{colorclock}{rgb}{0.4, 0.4, 1}
	\definecolor{colordisc}{rgb}{1, 0, 1}
	\definecolor{colorloc}{rgb}{0.4, 0.4, 0.65}
	\definecolor{colorparam}{rgb}{1, 0.6, 0.0}

\newcommand{\styleact}[1]{\ensuremath{\textcolor{coloract}{\mathrm{#1}}}}
\newcommand{\styleclock}[1]{\ensuremath{\textcolor{colorclock}{#1}}}

\newcommand{\styleloc}[1]{\ensuremath{\mathrm{#1}}}
\newcommand{\styleparam}[1]{\ensuremath{\textcolor{colorparam}{\mathrm{#1}}}}
\newcommand{\stylePTA}[1]{\ensuremath{\mathsf{#1}}}

\newcommand{\offsetTone}{\styleparam{offsetT1}}
\newcommand{\offsetTtwo}{\styleparam{offsetT2}}
\newcommand{\offsetTthree}{\styleparam{offsetT3}}
\newcommand{\deadlineTone}{\styleparam{deadlineT1}}
\newcommand{\deadlineTtwo}{\styleparam{deadlineT2}}
\newcommand{\deadlineTthree}{\styleparam{deadlineT3}}

\ifdefined \VersionWithComments
 	\definecolor{colorok}{RGB}{80,80,150}
\else
	\definecolor{colorok}{RGB}{0,0,0}
\fi

\newcommand{\eg}{\textcolor{colorok}{e.\,g.,}\xspace}
\newcommand{\ie}{\textcolor{colorok}{i.\,e.,}\xspace}

\newcommand{\wrt}{\textcolor{colorok}{w.r.t.}\xspace}

\title{Parametric schedulability analysis of a launcher flight control system under reactivity constraints
\thanks{%
\LongVersion{%
	This manuscript is the extended version of the manuscript of the same name published in the proceedings of the 19th International Conference on Application of Concurrency to System Design (ACSD 2019).
	The final version is available at \url{https://ieeexplore.ieee.org/}.
}
	This work was supported by the Paris Île-de-France Region (project DIM RFSI ASTREI).
	Étienne André is partially supported by
	the ANR national research program PACS (ANR-14-CE28-0002),
	and
	the
	ERATO HASUO Metamathematics for Systems Design Project (No.\ JPMJER1603), JST.
}
}

\author{\IEEEauthorblockN{Étienne André}
\IEEEauthorblockA{\textit{Université Paris 13, LIPN, CNRS, UMR 7030, F-93430, Villetaneuse, France}\\
\textit{JFLI, CNRS / NII, Tokyo, Japan}
}
\and
\IEEEauthorblockN{Emmanuel Coquard}
\IEEEauthorblockA{\textit{ArianeGroup SAS} \\
Les Mureaux, France \\
}
\and
\IEEEauthorblockN{Laurent Fribourg}
\IEEEauthorblockA{\textit{LSV, ENS Paris-Saclay \& CNRS \& INRIA, France} \\
Cachan, France \\
}
\and
\IEEEauthorblockN{Jawher Jerray}
\IEEEauthorblockA{\textit{Université Paris 13, LIPN, CNRS, UMR 7030} 
\\
Villetaneuse, France \\
}
\and
\IEEEauthorblockN{David Lesens}
\IEEEauthorblockA{\textit{ArianeGroup SAS} \\
Les Mureaux, France \\
}
}

\begin{document}

\pagestyle{plain}

\maketitle

\thispagestyle{plain}

\ifdefined \VersionWithComments
	\textcolor{red}{\textbf{This is the version with comments. To disable comments, comment out line~3 in the \LaTeX{} source.}}
\fi

\begin{abstract}
	The next generation of space systems will have to achieve more and more complex missions.
	In order to master the development cost and duration of such systems, an alternative to a manual design is to automatically synthesize the main parameters of the system.
	In this paper, we present an approach on the specific case of the scheduling of the flight control of a space launcher.
	The approach requires two successive steps: (1) the formalization of the problem to be solved in a parametric formal model and (2) the synthesis of the model parameters with a tool.
	We first describe the problematic of the scheduling of a launcher flight control, then we show how this problematic can be formalized with parametric stopwatch automata; we then present the results computed by \imitator{}.
	We compare the results to the ones obtained by other tools classically used in scheduling.
\end{abstract}



\section{Introduction}\label{section:introduction}

Real-time systems combine concurrent behaviors with hard timing constraints.
An out-of-date reply is often considered as invalid even if its content is correct.
For \emph{critical} real-time systems, if a time constraint is violated, then the consequences can be disastrous.
Thus, a formal verification phase 
is essential in order to statically guarantee that all the tasks will be executed in their allocated time, and that the system will return results within the times guaranteed by the specification.

Assessing the absence of timing constraints violations is even more important when the system can be hardly controlled once launched.
This is especially true in the aerospace area, where a system can only very hardly be modified or even rebooted after launching.

While verifying a real-time system is already a notoriously difficult task, we tackle here the harder problem of synthesis, \ie{} to automatically synthesize a part of the system so that it meets its specification.
%
The next generation of space systems will have to achieve more and more complex missions.
In order to master the development cost and duration of such systems, an alternative to a manual design is to automatically synthesize the main parameters of the system.

\paragraph{Contribution}
In this paper, we address the specific case of the scheduling of the flight control of a space launcher.
Our approach requires two successive steps:
\begin{ienumeration}
	\item the formalization of the problem to be solved in a parametric formal model and,
	\item the synthesis of the model parameters with a tool.
\end{ienumeration}
We first describe the problematic of the scheduling of a launcher flight control, then we formalize this problematic with parametric stopwatch automata, and third we present the results computed by the \imitator{} tool.
We compare our results with the ones obtained by other tools classically used in scheduling.
A key aspect is the verification and synthesis under some reactivity constraints: the time from a data generation to its output must always be less to a threshold.
The solution we propose is compositional.

We propose here a solution to the problems using an extension of parametric timed automata (PTAs), an extension of finite state automata with clocks and parameters~\cite{AHV93}.
PTAs are notoriously undecidable (see~\cite{Andre19STTT} for a survey), despite some decidable subclasses (\eg{} \cite{HRSV02,BlT09,ALime17,ALR18FORMATS}), notably in the field of scheduling real-time systems~\cite{CPR08,Andre17FMICS}.
In spite of these undecidability results, we show that this formalism is handful for solving concrete problems---such as the one considered here.

\paragraph{Outline}
After discussing related works in \cref{section:related},
\cref{section:problem} presents the problem we aim at solving.
\cref{section:PSA} recalls parametric stopwatch automata.
\cref{section:specification,section:reactivities} expose our modeling, while \cref{section:experiments} gives the results obtained.
\cref{section:comparison} makes a comparison with additional tools of the literature (solving only a part of the problem).
\cref{section:conclusion} concludes the paper.

\section{Related works}\label{section:related}

\paragraph{Scheduling}
A long line of works in the last four decades has been devoted to the problem of scheduling analysis of real-time systems with various flavors.
%
Several analytical methods were proposed to study the schedulability for a particular situation.
Such analytical methods need to be tuned for each precise setting (uniprocessor or multiprocessor, scheduling policy, absence or presence of offsets, jitters, etc.).
Most of them do not cope well with uncertainty.
For example, in~\cite{BB97}, three methods for the schedulability analysis with offsets are proposed.
In~\cite{BB04}, an efficient approach for testing schedulability for \RMS{} (rate monotonic) in the case of (uniprocessor) analysis is proposed, through a ``parameter'' (different from our timing parameters) to balance complexity versus acceptance ratio.

\paragraph{Scheduling with model checking}
Schedulability with model checking is a trend that started as early as the first works on timed model checking (\eg{} \cite{WME92,AHV93,AD94,YMW97,CC99}), and grew larger since the early 2000s.
On the negative side, the cost of state space explosion often prevents to verify very large real-time systems.
On the positive side, they allow for more freedom, and can model almost any system with arbitrarily complex constraints;
in addition, despite the cost of state space explosion, they can be used to verify small to medium-size systems for which no other method is known to apply.

A natural model to perform schedulability analysis is (extensions of) timed automata (TAs)~\cite{AD94}.
In \ShortVersion{\cite{AAM06}}\LongVersion{\cite{AM01,AM02,AAM06}}, \LongVersion{(acyclic) }TAs are used to solve job-shop and scheduling problems.
This allows to model naturally more complex systems which are not captured so easily in traditional models of operation research.

In \cite{NWY99,FKPY07,Andre17FMICS}, task automata are proposed as a formalism extending TAs to ease the modeling (and the verification) of uniprocessor real-time systems: in some cases, the schedulability problem of real-time systems is transformed into a reachability problem for standard TAs and it is thus decidable.
This allows to apply model-checking tools for TAs to schedulability analysis with several types of tasks and most types of scheduler.\ea{remove if space needed?}
 
In \cite{SLSFM14}, hierarchical scheduling systems are encoded using linear hybrid automata, a model that generalizes TAs.
This approach outperforms analytical methods in terms of resource utilization.
In~\cite{SL14}, linear hybrid automata are used to perform schedulability analysis for multiprocessor systems under a global fixed priority scheduler: this method is more scalable than existing exact methods, and shows that analytical methods are pessimistic.

In~\cite{FLSC16}, a schedulability analysis method is introduced using the model of \emph{timed regular task automata} (using under-approximated WCETs) and then using nested timed automata (which is exact).

The problem we solve here shares similarities with analyses done in~\cite{FBGLP10,MLRNSPPH10}.
An important difference between \cite{FBGLP10,MLRNSPPH10} and our case study comes from the fact that, here, there are two distinct notions of ``thread'' and ``processing'', while in~\cite{FBGLP10,MLRNSPPH10} there was only one notion called ``task''.
Most importantly, none of these works consider timing parameters.
%
%

\paragraph{Scheduling with parameters}
When some of the design parameters are unknown or imprecise, the analysis becomes much harder.
Model checking with parameters can help to address this.
In~\cite{CPR08}, PTAs are used to encode real-time systems so as to perform parametric schedulability analysis.
A subclass (with bounded offsets, parametric WCETs but constants deadlines and periods) is exhibited that gives exact results.
In contrast, our work allows for parameterized deadlines; in addition, reactivities are not considered in~\cite{CPR08}.

In~\cite{FLMS12}, we performed robust schedulability analysis on an industrial case study, using the inverse method for PTAs~\cite{ACEF09} implemented in \imitator{}.
While the goal is in essence similar to the one in this manuscript, the system differs: \cite{FLMS12} considers multiprocessor, and preemption can only be done at fixed instants, which therefore resembles more Round Robin than real \FPS{}.
In~\cite{SSLAF13}, we showed that PTAs-based methods are significantly more complete than existing analytical methods to handle uncertainty.
In~\cite{SAL15}, we solved an industrial challenge by Thales using \imitator{}.

In \cite{LPPR13}, the analysis is not strictly parametric, but concrete values are iterated so as to perform a cartography of the schedulability regions.
However, the resulting analysis of the system is incomplete.

\LongVersion{%
In \cite{BHJL16}, timed automata are ``extended'' with multi-level clocks, of which exactly one at a time is active.
The model enjoys decidability results, even when extended with polynomials and parameters, but it remains unclear whether concrete classes of real-time systems can actually be modeled.

Finally, \textsc{Roméo}~\cite{LRST09} also allows for parametric schedulability analysis using parametric time Petri nets~\cite{TLR09}.
}

\todo{\cite{MLRNSPPH10}}
\todo{\cite{FBGLP10}}

\section{Description of the system and problem}\label{section:problem}

The flight control of a space launcher is classically composed of three algorithms:
\begin{itemize}
\item The \emph{navigation} computes the current position of the launcher from the sensor's measurement (such as inertial sensors);
\item The \emph{guidance} computes the most optimized trajectory from the launch pad to the payload release location;
\item The \emph{control} orientates the thruster to follow the computed trajectory.
\end{itemize}
Due to the natural instability of a space launcher, strict real-time requirements have to be respected by the implementation of the flight control: frequency of each algorithm and reactivity between the sensor's measurement acquisition and the thruster's command's sending.

The case study described in this paper is a simplified version of a flight control composed of a navigation, a guidance, a control and a monitoring algorithms (also called \emph{processings}).
Each processing has a name and a required period.
A processing can potentially read data from the avionics bus (``in'' data) and / or write data to the same avionics bus (``out'' data).
\cref{figure:example_control_system} shows an example of such system (all the numerical data provided in this paper are only examples which do not correspond necessarily to an actual system).

\ShortVersion{
\begin{figure}[tb]
  \centering
  
\begin{lstlisting}[style=ariane,basicstyle=\scriptsize]
processing Navigation (Meas     : in)  is period (5ms);
processing Guidance                    is period (60ms);
processing Control    (Cmd      : out) is period (10ms);
processing Monitoring (Safeguard: out) is period (20ms);
\end{lstlisting}
  \caption{An example of a flight control system}
  \label{figure:example_control_system}
\end{figure}
}
\LongVersion{
\begin{figure*}[tb]
	\centering
\begin{lstlisting}[style=ariane]
processing Navigation (Meas     : in)  is period (5ms);  end;
processing Guidance                    is period (60ms); end;
processing Control    (Cmd      : out) is period (10ms); end;
processing Monitoring (Safeguard: out) is period (20ms); end;
\end{lstlisting}
	\caption{An example of a flight control system}
	\label{figure:example_control_system}
\end{figure*}
}

\subsection{Threads and deterministic communications}\label{ss:threads}

Processings are allocated on \emph{threads} run by the processor.
In our setting, all the thread's periods are \emph{harmonic}, \ie{} a thread period is a multiple of the period of the thread just smaller (they pairwise divide each other).


In addition, in order to ensure the determinism of the scheduling (which facilitates the verification of the system), the threads work in a synchronous manner:
\begin{itemize}
	\item The inputs of a thread are read at its start (no inputs are read during the execution of the thread)
	\item The outputs of a thread are provided at its deadline (no outputs are provided during the execution of the thread)
\end{itemize}

\begin{figure}[htb]
	\centering
  \includegraphics[width=\columnwidth]{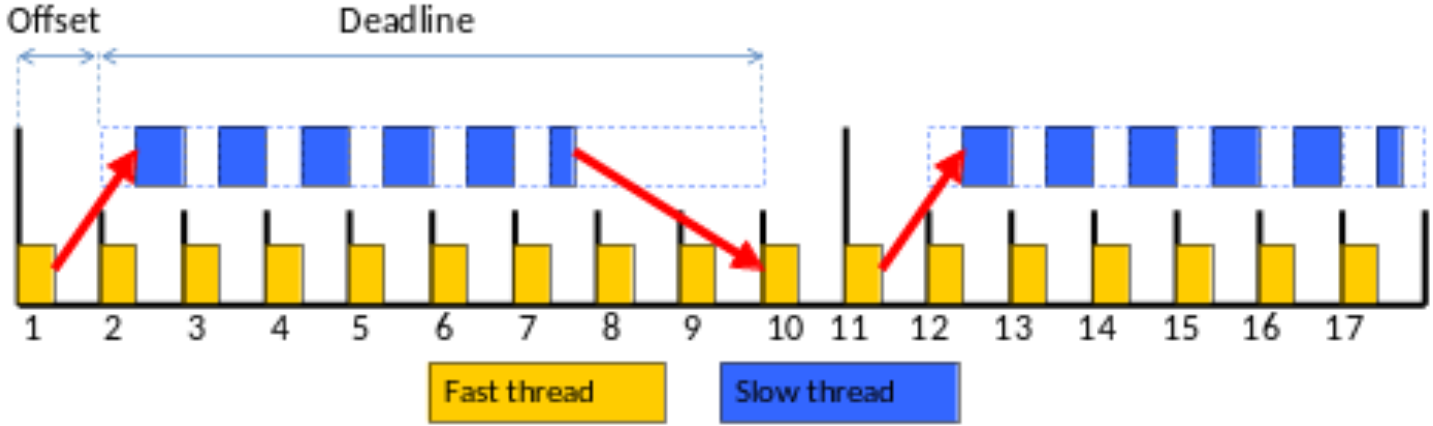}
  \caption{The communication between threads}
  \label{fig:communication_threads}
\end{figure}

\reviewAdd{\cref{fig:communication_threads}} shows the way data are exchanged between two threads.
The fast thread (in yellow) has a period of~1. This period defines the time granularity of the system (this implies that the offset of the fast thread is 0 and that its deadline is~1).
On this example, the slow thread (in blue) has an offset of~1 (its start is delayed of 1 cycle compared to the start of the fast thread) and a deadline of~8.
The communication between the fast thread and the slow one is performed immediately at the end of the first execution of the fast thread. In order to ensure the determinism and taking into the priority between the threads, the communication between the slow thread and the fast thread is performed at the deadline of the slow thread, \ie{} at the end of the cycle 9 (offset + deadline).

\subsection{Reactivities}\label{ss:reactivities}

To ensure the controllability of the launcher, a \emph{reactivity} is required between a data read from the avionics bus (a measurement) and a data written to the avionics bus (a command).
Several paths are potentially possible between a read data and a written data.
\cref{figure:reactivities} shows an example of such reactivities.

\begin{figure}[htbp!]
  \centering
  
\begin{lstlisting}[style=ariane]
reactivity Meas -> Navigation -> Guidance  -> Control -> Cmd       is 150ms;
reactivity Meas -> Navigation -> Control              -> Cmd       is 15ms;
reactivity Meas -> Navigation -> Monitoring           -> Safeguard is 55ms;
\end{lstlisting}
  \caption{Some typical reactivities}
  \label{figure:reactivities}
\end{figure}

\begin{figure*}[htbp!]
  \centering
  
  \includegraphics[width=.8\textwidth]{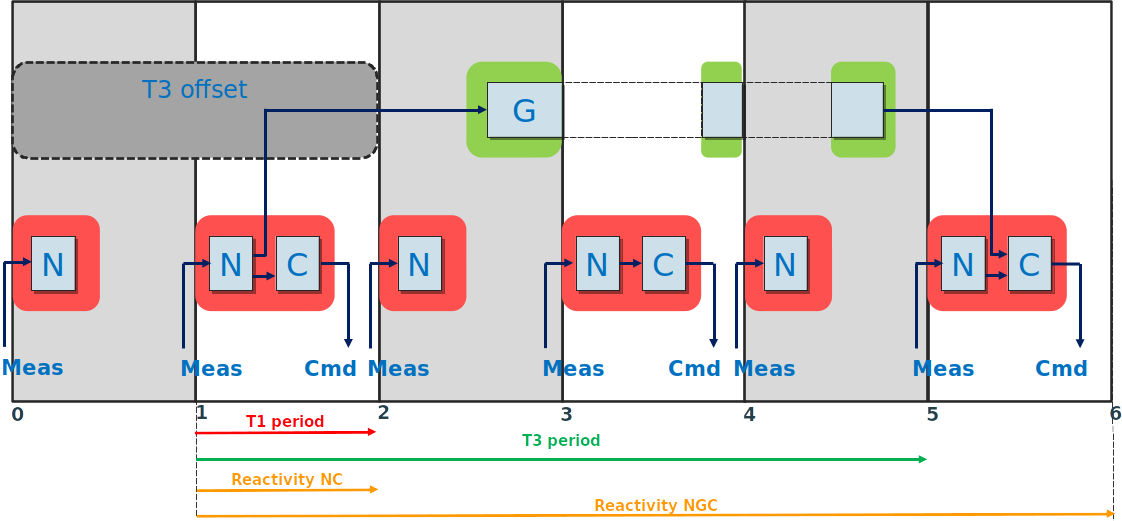}
  \caption{Determinism and reactivities}
  \label{figure:example-reactivities}
\end{figure*}

We want to solve the scheduling problem of periodic processing under precedence and \textit{reactivity constraints}, as in~\cite{FBGLP10}.
Reactivities too must follow the deterministic communication model from \cref{ss:threads}.
Consider the reactivity ``Meas $\rightarrow$ Navigation $\rightarrow$ Guidance $\rightarrow$ Control $\rightarrow$ Cmd'' depicted in \reviewAdd{\cref{figure:example-reactivities}} (the values of periods and WCETs are not necessarily the ones given in our case study).
Due to the data being communicated at the end of each thread only, the Guidance processing (marked with ``G'' in green) does not receive the data from the third execution of the Navigation processing (marked with ``N'' in red), as the data of the third Navigation will be sent at the end of the thread T1 period, but from the second execution of Navigation.
Therefore, in \reviewAdd{\cref{figure:example-reactivities}}, the only path of interest is the path of the data starting from the second execution of Meas, going to the second execution of Navigation, then going to the (only) execution of Guidance, and then finishing in the third execution of Control, before being written to the third occurrence of Cmd.
Also note that the data output by the first execution of Navigation are successfully sent to T2 at the end of the first period of~T1, but will be overwritten by the second occurrence of Navigation, and are therefore not of interest for the computation of reactivities.

\subsection{Processings and assignment into threads}

A WCET (worst case execution time) is measured or computed for each processing.
An example is given in \cref{figure:example_wcet}.

\begin{figure}[htbp!]
  \centering
  
\begin{lstlisting}[style=ariane]
processing wcet Navigation (1ms);
processing wcet Guidance   (15ms);
processing wcet Control    (3ms);
processing wcet Monitoring (5ms);
\end{lstlisting}
  \caption{Example of Worst Case Execution Times}
  \label{figure:example_wcet}
\end{figure}

An important problem is to find a proper assignment of the processings into threads, with their respective periods.
A solution to this problem consists on a set of cyclic threads on which the processings are deployed.
In our setting, these threads 
are scheduled with a preemptive and fixed priority policy (\FPS{}). 
A thread has a name and is defined by the following data:
\begin{itemize}
	\item a rational-valued period;
	\item a rational-valued offset (with offset $\leq$ period), \ie{} the time from the system start until the first periodic activation;
	\item a rational-valued (relative) deadline (with deadline $\leq$ period), \ie{} the time after each thread activation within which all processings of the current thread should be completed;
	\item a rational-valued major frame (or ``MAF''). A MAF defines the duration of a pattern of processing activation;
	\item a set of processings deployed in the thread.
		Different processings may be executed at each cycle. However, after a MAF duration, the same pattern of processings is repeated.
\end{itemize}

%

\subsection{Formalization}

A real-time system $\mathcal{S} = \{{\cal T}, {\cal P}, {\cal R}\}$ is viewed here as a set of \emph{threads} ${\cal T} = \{T_{1} , T_{2} , \cdots , T_{n} \}$, a set of \emph{processings} ${\cal P}= \{P_{1} , P_{2} , \cdots , P_{m} \}$ and a set of \emph{reactivities} ${\cal R}= \{R_{1} , R_{2} , \cdots, R_{q} \}$.
\LongVersion{A thread $T_{i}$ generates a possibly infinite stream of processings $P_{1} , P_{2}, \cdots$.}

A thread $T_{i}$ is periodic, and characterized by a 5-tuple $(PT_{i} , OT_{i} , DT_{i}, MAF_{i}, {\cal P}_{i})$, where $PT_{i}$ corresponds to the period, $OT_{i}$ to the offset, $DT_{i}$ to the deadline, $MAF_{i}$ defines the duration of a pattern of processing activation $P_{i,j}$, and ${\cal P}_{i}$ defines a subset of processings of ${\cal P}$ allocated to $T_{i}$.

A processing $P_i$ is characterized by two values $WCET_{i}$ and~$PP_{i}$.
When a processing is activated, it executes for at most time $WCET_{i}$, and has to terminate within the relative period $PP_{i}$.

\begin{figure}[htbp!]
  \centering
  
  \includegraphics[width=.5\textwidth]{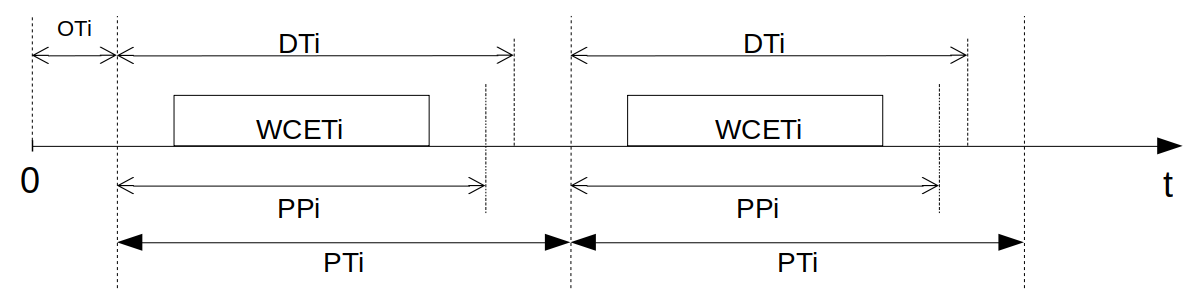}
  \caption{Real-time characteristics of the system}
  \label{figure:Real-time-characteristics-of-system}
\end{figure}

These definitions are illustrated in \cref{figure:Real-time-characteristics-of-system}.

A reactivity is of the form $R_{i} = ((P_{i,1} \rightarrow P_{i,2} \rightarrow \dots \rightarrow P_{i,k}),DR_{i} )$ where $(P_{i,1} \rightarrow P_{i,2} \rightarrow \dots \rightarrow P_{i,k})$ denotes a precedence constraint of ${\cal P}$, and $DR_{i}$ is the maximum \textit{time of reactivity} for $R_{i}$: the end of the thread containing the last processing $P_{i,k}$ of the precedence sequence has to be completed before the deadline $DR_{i}$.


\begin{definition}\label{definition:schedulability}
	A system $\mathcal{S}$ is \textit{schedulable} if
	\begin{itemize}
		\item $\forall$ $T_{i}$ $\in$ ${\cal T}$, the end of $T_{i}$ occurs before $DT_{i}$.
		\item $\forall$ $R_{i}$ $\in$ ${\cal R}$, the end of thread containing the last processing $P_{i,k}$ of $R_{i}$ occurs before~$DR_{i}$.
	\end{itemize}
\end{definition}

\LongVersion{
\subsection{Description of the case study}

We give here the values for the case study of interest.

\LongVersion{\subsubsection{Processings}}

The processings $P_{i}$ considered corresponds to the lists: 
\begin{itemize}
\item $P_{1}$(Navigation) = $(WCET_{1} , PP_{1}) = (1, 5) $

\item $P_{2}$(Control) = $(WCET_{2} , PP_{2}) = (3, 10) $

\item $P_{3}$(Monitoring) = $(WCET_{3} , PP_{3}) = (5, 20) $

\item $P_{4}$(Guidance) = $(WCET_{4} , PP_{4}) = (15, 60) $
\end{itemize}

%
%
%

\LongVersion{\subsubsection{Threads}}

The threads $T_{i}$ considered corresponds to the lists: 
\begin{itemize}
\item $T_{1}$ = $(PT_{1} , OT_{1} , DT_{1}, MAF_{1}, {\cal P}_{1}) = (5, OT_{1}, DT_{1}, 10, \{P_{1}, P_{2}\}) $

\item $T_{2}$ = $(PT_{2} , OT_{2} , DT_{2}, MAF_{2}, {\cal P}_{2}) = (20, OT_{2}, DT_{2}, 20, \{P_{3}\})$

\item $T_{3}$ = $(PT_{3} , OT_{3} , DT_{3}, MAF_{3}, {\cal P}_{3}) = (60, OT_{3}, DT_{3}, 60, \{P_{4}\})$

\end{itemize}

\LongVersion{\subsubsection{Reactivities}}\label{subsubsection:reactivities}
The reactivities (end-to-end flow) $R_{i}$ considered correspond to the lists: 
\begin{itemize}
\item $R_{1} = ((P_{1,1},P_{1,4},P_{1,2}),DR_{1})$ where $(P_{1,1} \rightarrow P_{1,4} \rightarrow P_{1,2})$ = (Navigation, Guidance, Control) and $DR_{1}$ = 150.
\item $R_{2} = ((P_{2,1},P_{2,2}),DR_{2})$ where $(P_{2,1} \rightarrow P_{2,2})$ = (Navigation, Control) and $DR_{2}$ = 15.
\item $R_{3} = ((P_{3,1},P_{3,3}),DR_{3})$ where $(P_{3,1} \rightarrow P_{3,3})$ = (Navigation, Monitoring) and $DR_{3}$ = 55.
\end{itemize}
}

\subsection{Objectives}

\begin{figure*}
	\centering
	\begin{subfigure}[b]{.36\textwidth}
\begin{lstlisting}[style=ariane]
thread T1 is
  period     (5ms);
  offset     (0ms);
  deadline   (5ms);
  maf        (10ms);
  processing (when 0 => (Navigation);
              when 1 => (Navigation; Control));end;
\end{lstlisting}
		\caption{T1}
		\label{fig:T1}
	\end{subfigure}
	\hspace{1em}
	\begin{subfigure}[b]{0.25\textwidth}
\begin{lstlisting}[style=ariane]
thread T2 is
  period     (20ms);
  offset     (0ms);
  deadline   (20ms);
  maf        (20ms);
  processing(Monitoring);
end;
\end{lstlisting}
		\caption{T2}
		\label{fig:T2}
	\end{subfigure}
	\hspace{1em}
	\begin{subfigure}[b]{0.25\textwidth}
\begin{lstlisting}[style=ariane]
thread T3 is
  period     (60ms);
  offset     (0ms);
  deadline   (60ms);
  maf        (60ms);
  processing (Guidance);
end;
\end{lstlisting}
		\caption{T3}
		\label{fig:T3}
	\end{subfigure}
	\caption{A typical solution of the flight control scheduling problem}
	\label{figure:typical_solution}
\end{figure*}

In order to simplify the scheduling problem, we have considered in this paper a pre-allocation of processings on threads, as specified in \cref{figure:typical_solution}: that is, Navigation and Control are allocated on T1, while Monitoring and Guidance are allocated on T2 and~T3, respectively.
In addition, Navigation is executed at every period of T1, while Control is executed (after Navigation) on \emph{odd} cycles only; this is denoted by the \texttt{when 1} syntax in \cref{fig:T1}.
The offsets and deadlines of each thread are unknown; that is, the values in \cref{figure:typical_solution} are not part of the input of the problem.
The flight control scheduling problem consists thus in computing the offsets and deadlines of each thread in order to fulfill the required reactivities.

Let us summarize the problems we address in this paper.
Our problems take as input:
\begin{enumerate}
	\item a flight control system \ie{} a list of processings with their period, and their input or output data (for example \cref{figure:example_control_system});
	\item a set of reactivities (for example \cref{figure:reactivities});
	\item a set of WCET for the processings (for example \cref{figure:example_wcet});
	\item an allocation of processings on threads (for example \cref{figure:typical_solution}).
\end{enumerate}

A solution to this problem is a set of threads with their period, offset, deadline and MAF (the MAF can in fact be obtained immediately from the period).
A solution is \emph{correct} if the set of reactivities is schedulable (as in \cref{definition:schedulability}).

The first problem is to formally \emph{verify} a solution to the problem:

\smallskip

\defProblem
	{scheduling verification}
	{a flight control system, a set of reactivities, a set of WCET, an allocation of processings on threads, and a solution}
	{formally verify that $\mathcal{S}$ is schedulable.}

\smallskip

The second problem is to \emph{synthesize} solutions to the problem:

\smallskip

\defProblem
	{scheduling synthesis}
	{a flight control system, a set of reactivities, a set of WCET, an allocation of processings on threads}
	{exhibit correct solutions.}

Recall that our synthesis problem still considers as input the periods, therefore offsets and deadlines are the main results of interest.

\section{Parametric stopwatch automata}\label{section:PSA}


\paragraph{Clocks, parameters, constraints}
We assume a set~$\Clock = \{ \clock_1, \dots, \clock_\ClockCard \} $ of \emph{clocks}, \ie{} real-valued variables that evolve at the same rate.
\LongVersion{A clock valuation is\LongVersion{ a function} $\clockval : \Clock \rightarrow \Rgeqzero$. 
We write $\ClocksZero$ for the clock valuation assigning $0$ to all clocks.
Given $\resets \subseteq \Clock$, we define the \emph{reset} of a valuation~$\clockval$, denoted by $\reset{\clockval}{\resets}$, as follows:
$\reset{\clockval}{\resets}(\clock) = 0$ if $\clock \in \resets$, and
$\reset{\clockval}{\resets}(\clock)=\clockval(\clock)$ otherwise.
Given a valuation~$\clockval$, $d \in \grandrplus$ and $\Clock' \subseteq \Clock$, we define the \emph{time-elapsing of~$\clockval$ by~$d$ except for clocks in~$\Clock'$}, denoted by $\timelapseStop{\clockval}{d}{\Clock'}$, as the clock valuation such that
\[\timelapseStop{\clockval}{d}{\Clock'}(\clock) = \begin{cases} 
      \clockval(\clock) & \text{if } \clock \in \Clock' \\
      \clockval(\clock) + d & \text{otherwise}\\
   \end{cases}
\]
}

We assume a set~$\Param = \{ \param_1, \dots, \param_\ParamCard \} $ of
\emph{parameters}\LongVersion{, \ie{} unknown constants}.  A parameter {\em
valuation} $\pval$ is\LongVersion{ a function} $\pval : \Param \rightarrow
\grandqplus$.
We denote ${\compOp} \in \{<, \leq, =, \geq, >\}$.
A guard~$\guard$ is a
constraint over $\Clock \cup \Param$ defined by a conjunction of inequalities
of the form $\clock \compOp d$ or $\clock \compOp \param$, with
$\clock\in\Clock$, $d \in \grandn$ and $\param \in \Param$.
\LongVersion{Given a guard
$\guard$, we write~$\clockval\models\pval(\guard)$ if the expression obtained
by replacing in~$\guard$ each~$\clock\in\Clock$ by~$\clockval(\clock)$ and
each~$\param\in\Param$ by~$\pval(\param)$ evaluates to true.  
}

\paragraph{Parametric stopwatch automata}
Parametric timed automata (PTA) extend timed automata with parameters within guards and invariants in place of integer constants~\cite{AHV93}.
For many real-time systems, especially when they are subject to preemptive scheduling, parametric timed automata are not sufficiently expressive.
As a result, we will use here an extension of PTA with stopwatches \cite{CL00}, namely parametric stopwatch automata~\cite{SSLAF13}.

\begin{definition}[PSA]\label{def:PSA}
	A \emph{parametric stopwatch automaton} (PSA) $\A$ is a tuple \mbox{$\A = (\Actions, \Loc, \locinit, \Clock, \Param, \invariant, \stopFunction, \Edges)$}, where:
    \begin{itemize}
		\item $\Actions$ is a finite set of actions,
		\item $\Loc$ is a finite set of locations,
		\LongVersion{\item }$\locinit \in \Loc$ is the initial location,
		\item $\Clock$ is a finite set of clocks,
		\item $\Param$ is a finite set of parameters,
        \item $\invariant$ is the invariant, assigning to every $\loc\in \Loc$ a guard $\invariant(\loc)$,
        \item $\stopFunction$ is the stop function $\stopFunction : \loc \rightarrow 2^\Clock$, assigning to every $\loc \in \Loc$ a set of stopped clocks,
		\item $\Edges$ is a finite set of edges  $\edge = (\loc,\guard,\action,\resets,\loc')$
		where~$\loc,\loc'\in \Loc$ are the source and target locations,
			$\guard$ is a guard,
			$\action \in \Actions$,
			and
			$\resets\subseteq \Clock$ is the set of clocks to be reset.
    \end{itemize}
\end{definition}

Stopwatch automata can be composed as usual using parallel composition on synchronized actions.
Note that our clocks are \emph{shared} by default, \ie{} a same clock (\ie{} with the same name) can be read, stopped or reset in several automata.
The same applies to parameters.

Given a parameter valuation~$\pval$ and PSA~$\A$, we denote by
$\valuate{\A}{\pval}$ the non-parametric structure where all occurrences of a
parameter~$\param\in\Param$ have been replaced by~$\pval(\param)$.
Any structure $\valuate{\A}{\pval}$ is also a \emph{stopwatch automaton}~\cite{CL00}.
If $\stopFunction(\loc) = \emptyset$ for all $\loc \in \Loc$, then by assuming a rescaling of the constants (multiplying all constants in $\valuate{\A}{\pval}$ by their least common denominator), we obtain an equivalent (integer-valued) TA\LongVersion{, as defined in}~\cite{AD94}.

\LongVersion{Let us now recall the concrete semantics of stopwatch automata.

\begin{definition}
	Given a PSA $\A = (\Actions, \Loc, \locinit, \Clock, \Param, \invariant, \stopFunction, \Edges)$,
	and a parameter valuation~\(\pval\),
	the semantics of $\valuate{\A}{\pval}$ is given by the timed transition system (TTS) $(\States, \sinit, \flecheRel)$, with
	\begin{itemize}
		\item $\States = \{ (\loc, \clockval) \in \Loc \times \Rgeqzero^\ClockCard \mid \clockval \models \valuate{\invariant(\loc)}{\pval} \}$, 
		\LongVersion{\item }$\sinit = (\locinit, \ClocksZero) $,
		\item  $\flecheRel$ consists of the discrete and (continuous) delay transition relations:
		\begin{ienumeration}
			\item discrete transitions: $(\loc,\clockval) \longueflecheRel{\edge} (\loc',\clockval')$, 
				if $(\loc, \clockval) , (\loc',\clockval') \in \States$, and there exists $\edge = (\loc,\guard,\action,\resets,\loc') \in \Edges$, such that $\clockval'= \reset{\clockval}{\resets}$, and $\clockval\models\pval(\guard$).
			\item delay transitions: $(\loc,\clockval) \longueflecheRel{d} (\loc, \timelapseStop{\clockval}{d}{\stopFunction(\loc)} )$, with $d \in \Rgeqzero$, if $\forall d' \in [0, d], (\loc, \timelapseStop{\clockval}{d'}{\stopFunction(\loc)}) \in \States$.\ea{Laurent: j'ai un peu improvisé pour les stopwatches ici, n'hésite pas à jeter un coup d'œil}
		\end{ienumeration}
	\end{itemize}
\end{definition}

}

\ShortVersion{The semantics of PSA is recalled in~\cite{ACFJL19report}.}

%
%
%

\section{Specifying the system}\label{section:specification}

Since the seminal work of Liu and Layland\LongVersion{ in~\cite{LL73}}, plenty of methods and tools have been designed to verify real-time systems.
However, while some aspects are reasonably easy (\FPS{}, no mixed-criticality), the problem we address here is not typical for several reasons:
\begin{itemize}
	\item offsets may be non-null;
	\item threads periods are harmonic; 
	\item the executed processings may differ depending on the cycle;
	\item the reactivities must always be met, and therefore define new, non-classical timing constraints;
	\item and, perhaps most importantly, the admissible values for deadlines and offsets may not be known. Only the global end to end reactivity is specified.
\end{itemize}

As a consequence, we choose to follow a \emph{model checking} based method.
Model checking is known for being more expressive than analytical methods, at the cost of performance or even decidability.
We show here that, although we use an undecidable formalism, we do get exact results for the instance of the problem we considered.

We present in the remainder of this section our modeling of the verification and the synthesis problem using parametric stopwatch automata.
This formalism has several advantages.
First, it is handful to model concurrent aspects of the system (different threads and processings running concurrently).
Second, stopwatches can be used to model preemption.
Third, parameters can be used to model the unknown constants, and solve the synthesis problem.




\subsection{Architecture of the solution}

\paragraph{A modular solution}
To model the system, we use the concurrent structure of parametric stopwatch automata to allow for a modular solution: that is, each element (thread, processing, scheduling policy) and each constraint (reactivity) is defined by an automaton.
These automata are then composed by usual parallel composition on synchronization actions.

This makes our solution modular in the sense that, in the case of a modification in the system (\eg{} the scheduling policy), we can safely replace one automaton by another (\eg{} the \FPS{} scheduler automaton with another scheduler automaton) without impacting the rest of the system.

\paragraph{Encoding elements and constraints as automata}
We model each processing activation as an automaton.
These automata ensure that processings are activated periodically with their respective period, and initial offset.

In addition, we create one automaton for each thread: the purpose of these automata is to 
	ensure that the processings associated with each thread are executed at the right time.
In the case of our concrete problem, we assign both the Navigation and Control processings to the T1 thread, the monitoring process to~T2 and the guidance processing to~T3.

The reactivities also follow the concept of modularity.
That is, each reactivity is \emph{tested} using a single automaton.
By testing, we mean that a reactivity fails iff a special location is reached.
Therefore, ensuring validity of the reactivities is equivalent to unreachability of these special locations.

\LongVersion{Finally, we specify a scheduler automaton which provides the scheduling between the different threads (in the case of our problem, recall that the scheduling policy is fixed priority scheduling (\FPS{}).}

We give more details on each of these automata in the following.

\todo{dire quelque chose sur les actions de synchronisation ; parler d'activation / start / finish}

\subsection{Modeling periodic processing activations}

Each processing is defined by a period and an offset.
To model the periodicity of the processings, we create one automaton for each processing activation.
This automaton simply performs the activations in a periodic manner.
Activations are modeled by a synchronization action, used to communicate with other automata (typically the thread automaton).

In addition, the period processing activation automaton detects whether a processing missed its (implicit) deadline equal to its period;
	that is, we assume that a processing that has not finished by the next period is a situation to be avoided.

Each such automaton features a single clock.

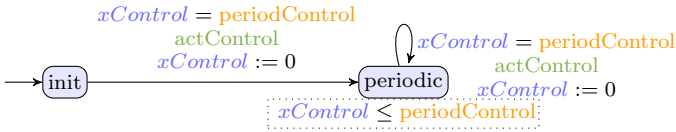
\begin{figure}
	\centering
\scriptsize
	\begin{tikzpicture}[scale=1.5, auto, ->, >=stealth']
 
		\node[location, initial] at (0,0) (l1) {\styleloc{init}};
 
		\node[location] at (3, 0) (l2) {\styleloc{periodic}};
		\node [invariant,below] at (l2.south) {\begin{tabular}{@{} c @{\ } c@{} }& $ \styleclock{xControl} \leq \styleparam{periodControl}$\\\end{tabular}};

		\path (l1) edge node[above]{\begin{tabular}{@{} c @{\ } c@{} }
		& $ \styleclock{xControl} = \styleparam{periodControl}$\\
		 & $\styleact{actControl}$\\
		 & $\styleclock{xControl}:=0$\\
		\end{tabular}} (l2);

		\path (l2) edge[loop above] node[below right]{\begin{tabular}{@{} c @{\ } c@{} }
		& $ \styleclock{xControl} = \styleparam{periodControl}$\\
		 & $\styleact{actControl}$\\
		 & $\styleclock{xControl}:=0$\\
		\end{tabular}} (l2);
	\end{tikzpicture}
%
	
	\caption{Automaton \stylePTA{periodicControl}}
	\label{figure:periodicControl}
\end{figure}

%

We present in \cref{figure:periodicControl} the example of the \stylePTA{periodicControl} automaton, modeling the periodic activation of the Control processing.
This controller contains a clock \styleclock{xControl} and one parameter \styleparam{periodControl}.
Note that the period \styleparam{periodControl} is known beforehand, and is therefore not strictly speaking a parameter, but that makes our solution both more generic and more readable (in \imitator{}, a parameter can be statically instantiated to a constant before running the analysis).


The initial location is \styleloc{init}: from then, the first occurrence of Control is immediately activated (action \styleact{actControl}), and the automaton enters the \styleloc{periodic} location.
Then, exactly every \styleparam{periodControl} time units (guard $\styleclock{xControl} = \styleparam{periodControl}$), another instance of Control is activated.

%
%

\subsection{Modeling threads}

\begin{figure*}
	\centering
	\scalebox{.9}{
	\begin{tikzpicture}[scale=2, auto, ->, >=stealth']
 
		\node[location, initial] at (0,0) (init) [align=center] {\styleloc{init}\\{\scriptsize stop $\{ \styleclock{xExecC} , \styleclock{xExecN} \}$}};
		\node [invariant, above] at (init.north) {\begin{tabular}{@{} c @{\ } c@{} }& $ \styleclock{xT1} \leq \styleparam{offsetT1}$\\\end{tabular}};
 
		\node[location] at (3, 0) (execNav) [align=center] {\styleloc{exec\_nav\_odd}\\{\scriptsize stop $\{ \styleclock{xExecC} \}$}};
		\node [invariant,above] at (execNav.north) {\begin{tabular}{@{} c @{\ } c@{} }& $ \styleclock{xT1} \leq \deadlineTone{}$ \\ $\land$ & $ \styleclock{xExecN} \leq \styleparam{WCETN}$ \end{tabular}};
 
		\node[location] at (3, -1.5) (execCon) [align=center] {\styleloc{exec\_control\_odd}\\{\scriptsize stop $\{ \styleclock{xExecN} \}$}};
		\node [invariant, below] at (execCon.south) {\begin{tabular}{@{} c @{\ } c@{} }& $ \styleclock{xT1} \leq \deadlineTone{}$ \\ $\land$& $ \styleclock{xExecC} \leq \styleparam{WCETC}$\end{tabular}};
 
		\node[location] at (0, -1.5) (idle) [align=center] {\styleloc{idle}\\{\scriptsize stop $\{ \styleclock{xExecC} , \styleclock{xExecN} \}$}};
		\node [invariant,below] at (idle.south) {\begin{tabular}{@{} c @{\ } c@{} }& $ \styleclock{xT1} \leq \styleparam{periodT1}$ \end{tabular}};
 
		\node[location, bad] at (6, -0.75) (missed) [align=center] {\styleloc{deadlineMissed}};

		\path (init) edge node[above]{\begin{tabular}{@{} c @{\ } c@{} }
		& $\styleclock{xT1} = \styleparam{offsetT1}$\\
		 & $\styleclock{xT1}:=0$\\
		\end{tabular}} (execNav);
 
		\path (execNav) edge node[right]{\begin{tabular}{@{} c @{\ } c@{} }
		& $ \styleclock{xExecN} = \styleparam{WCETN}$\\
		 & $\styleact{finishNavigation}$\\
		 & $\styleclock{xExecN}:=0$\\
		\end{tabular}} (execCon);
		
		\path (execCon) edge node[above]{\begin{tabular}{@{} c @{\ } c@{} }
		& $ \styleclock{xExecC} = \styleparam{WCETC}$\\
		 & $\styleact{finishControl}$\\
		 & $\styleclock{xExecC}:=0$\\
		\end{tabular}} (idle);

		\path (idle) edge[bend left] node[below left, xshift=-25, yshift=-10]{\begin{tabular}{@{} c @{\ } c@{} }
		& $ \styleclock{xT1} = \styleparam{periodT1}$\\
		 & $\styleclock{xT1}:=0$\\
		\end{tabular}} (execNav);
		
		\path (execNav) edge[bend left] node[above]{\begin{tabular}{@{} c @{\ } c@{} }
		& $ \styleclock{xT1} = \styleparam{deadlineT1}$\\
 		 $\land$ & $ (\styleclock{xExecC} < \styleact{WCETC}$\\
	 		& $ \lor \styleclock{xExecN} < \styleact{WCETN})$\\
		\end{tabular}} (missed);
		
		\path (execCon) edge[bend right] node[below]{\begin{tabular}{@{} c @{\ } c@{} }
		& $ \styleclock{xT1} = \styleparam{deadlineT1}$\\
 		 $\land$ & $ \styleclock{xExecC} < \styleact{WCETC}$\\
		\end{tabular}} (missed);
	\end{tikzpicture}
}
	
	\caption{Fragment of automaton \stylePTA{threadT1}}
	\label{figure:threadt1}
\end{figure*}
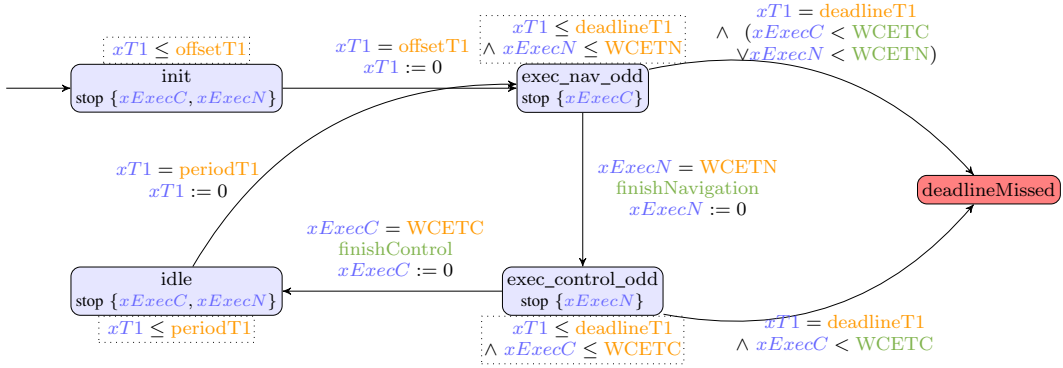

We create one automaton for each thread.
Each of these automata contains one clock for the thread (used to measure the thread period and offset), as well as one clock per processings assigned to the thread.
For example in \cref{figure:threadt1}, the thread automaton \stylePTA{threadT1} contains \styleclock{xT1} (the thread clock), as well as \styleclock{xExecControl} and \styleclock{xExecNavigation} (the clocks associated to the processings of~T1).
Parameters include the offset, period and deadline of the thread, but also the WCETs of the processings assigned to this thread.

The thread automaton is responsible for:
\begin{enumerate}
	\item encoding the initial thread offset, \ie{} starting the periodic thread activation only after the offset;
	\item performing the periodic thread activation;
	\item executing the processings associated with the thread;
	\item detecting the deadline misses.
\end{enumerate}

The clocks associated with the processings are used to measure the execution time of these processings: they are in fact stopped most of the time, except when the thread is actively executing the processing.
This is in contrast with the clocks associated with the processing activation automaton, that are never stopped, as they measure a period.
Then, a deadline miss occurs if the clock measuring the thread period reaches the deadline (recall that the deadline is less than or equal to the period, and therefore we can use the same clock), while the clock measuring a processing execution time is strictly less than its WCET.

We give in \cref{figure:threadt1} a fragment of the automaton \stylePTA{threadT1}.
We only give the odd cycle, as this is the most interesting; that is, we removed the fragment corresponding to the even cycle (only executing Navigation) between locations \styleloc{init} and \styleloc{exec\_nav\_odd} (and the transition from \styleloc{idle} should go to the removed \styleloc{exec\_nav\_even} location).
We also abbreviate some variable names to save space (\eg{} $\styleclock{xExecC}$ for $\styleclock{xExecControl}$  and  \styleclock{xExecN} for \styleclock{xExecNavigation} or \styleparam{WCETN} for \styleparam{WCETNavigation}).\LongVersion{

}%
First, the automaton waits for the offset: that is, it stays in \styleloc{init} exactly \styleparam{offsetT1} time units.
Then, it executes the first processing of the odd cycle, \ie{} Navigation: it stays in \styleloc{exec\_nav\_odd} until completion, \ie{} for \styleparam{WCETNavigation} time units.\footnote{%
	In the full model, we can allow for a best case execution time, in which case the duration is non-deterministically chosen in the interval $[\styleparam{BCETNavigation}, \styleparam{WCETNavigation}]$.
}
Note that this is the only location where \styleclock{xExecNavigation} is elapsing, \ie{} is not stopped, as it measures the execution time.
Then, upon completion of the Navigation processing, the automaton moves to \styleloc{exec\_control\_odd}, where Control is executed.
Upon completion, it moves to \styleloc{idle}, and waits until the clock $\styleclock{xT1}$ reaches its period.
Then, the cycle restarts and so on.

In addition, at any time, possible deadline misses are checked for.
A deadline miss occurs on an odd cycle while execution Navigation whenever $\styleclock{xT1} = \deadlineTone$ and either $\styleclock{xExecControl} < \styleparam{WCETControl}$ or $\styleclock{xExecNavigation} < \styleparam{WCETNavigation}$.\LongVersion{\footnote{%
	This encoding is not necessarily optimal.
	In fact, on odd cycles, as Navigation is executed first, and followed by Control, a deadline miss can be detected earlier, \ie{} if Navigation is still executed, but there is not enough time to finish the execution of Navigation and that of Control: that is, an optimized deadline miss condition could be $\styleclock{xT1} + \styleparam{WCETControl} = \deadlineTone$ and $\styleclock{xExecNavigation} < \styleparam{WCETNavigation}$.
	This optimization has not been implemented, so as to leave the model (relatively) simple and maintainable, but could be tested in the future.
}}
When executing Control, only the execution time of Control needs to be checked.

\LongVersion{%
\begin{remark}
Our model is in fact more complicated as, for sake of modularity, we make no assumption in the thread automaton on how the other automata behave, notably the processings activation automata.
	Therefore, we allow for processings to be activated at any time, which must be taken care of in the thread automaton.
\end{remark}
}

\subsection{Modeling the \FPS{} scheduler}

The \FPS{} scheduler is modeled using an additional automaton.
It reuses existing works from the literature (\eg{} \cite{FKPY07,SSLAF13}), and does not represent a significant original contribution.
We mainly reuse the scheduler encoding of \cite{SSLAF13}, which consists of an automaton synchronizing with the rest of the system on the start and end task synchronization actions as well as the task activation actions.
Whenever a new task is activated, the scheduler decides what to do depending on its current state and the respective priorities of the new and the executing tasks (if any).

Nevertheless, we had to modify this encoding due to the fact that existing scheduler automata simply schedule tasks: in the setting of our case study, the scheduler schedules both the threads and the threads' processings.
Among the various modifications, in case of preemption, our scheduler does not stop the clocks measuring the execution times of the preempted threads (because such clocks do not exist), but stop the clocks measuring the execution times of the \emph{processings belonging to the preempted threads}.

\begin{figure}
	\centering
 %

		\scalebox{.95}{
	\begin{tikzpicture}[scale=.8, xscale=2, auto, ->, >=stealth']
 
		\node[location, initial] at (0, 1) (idle) [align=center] {\styleloc{idle}  }; 
		\node [invariant, above left,align=center] at (idle.west) {};
 
		\node[location] at (2,2) (execT1) [align=center] {\styleloc{execT1}};
 
		\node[location] at (4, 1) (execT1waitT2) [align=center] {\styleloc{execT1waitT2} \\ stop$\{ \styleclock{xexecM} \}$};
 
		\node[location] at (2, 0) (execT2) [align=center] {\styleloc{execT2}};
 

		\path (idle) edge[bend angle=20,bend right] node{\begin{tabular}{@{} c @{\ } c@{} }
		 & $\styleact{actT1}$\\
		\end{tabular}} (execT1);

		\path (idle) edge[bend angle=20,bend left] node[]{\begin{tabular}{@{} c @{\ } c@{} }
		 & $\styleact{actT2}$\\
		\end{tabular}} (execT2);

		\path (execT1) edge[bend right] node[above left]{\begin{tabular}{@{} c @{\ } c@{} }
		 & $\styleact{finT1}$\\
		\end{tabular}} (idle);

		\path (execT1) edge node{\begin{tabular}{@{} c @{\ } c@{} }
		 & $\styleact{actT2}$\\
		\end{tabular}} (execT1waitT2);


		\path (execT1waitT2) edge[bend left] node[]{\begin{tabular}{@{} c @{\ } c@{} }
		 & $\styleact{finT1}$\\
		\end{tabular}} (execT2);



		\path (execT2) edge[bend left] node{\begin{tabular}{@{} c @{\ } c@{} }
		 & $\styleact{finT2}$\\
		\end{tabular}} (idle);

		\path (execT2) edge[bend left] node[above left]{\begin{tabular}{@{} c @{\ } c@{} }
		 & $\styleact{actT1}$\\
		\end{tabular}} (execT1waitT2);

 
	\end{tikzpicture}
		}
	
	\caption{Encoding the \FPS{} scheduler (simplified version)}
	\label{figure:scheduler}
\end{figure}
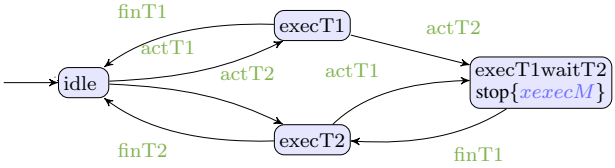

We give in \cref{figure:scheduler} an example of such a scheduler in a simplified version, with only two threads T1 and~T2\LongVersion{; the full scheduler is of course more complete}.
If any of the two threads get activated (\styleact{actT1} or \styleact{actT2}), the scheduler starts executing them.
If a second thread gets activated, the highest priority thread (T1) is executed, while T2 is put in waiting list (which is encoded in location \styleloc{execT1waitT2}).
This is the location responsible for stopping the clock of the (only) processing of~T2, \ie{} Monitoring (clock~$\styleclock{xexecM}$).
Only after T1 has completed (\styleact{finT1}), T2 can execute.
Our real scheduler is in fact significantly more complex as it has to cope with three threads, but also with special cases\LongVersion{ such as the activation of a new thread activation of~$T_i$ while executing a previous instance of~$T_i$, etc}.


%

\subsection{Reachability synthesis}\label{ss:reachability}

Finally, the system is schedulable if none of the ``bad'' locations (corresponding to deadline misses, \eg{} in the thread automata) is reachable.
If all parameters are valuated, the system is a TA, and schedulability reduces to reachability checking.
If some parameters are free (\ie{} the analysis is parametric), the set of valuations for which the system is schedulable exactly corresponds to the valuations for which these bad locations are unreachable, \ie{} the complement of the valuations set result of reachability synthesis.
This guarantees our method correctness.

\section{Compositional verification of reactivities}\label{section:reactivities}

An originality of our work---which among other reasons justifies our choice to use model checking
---is the encoding of \emph{reactivities}.
Indeed, our goal is to verify a system, or synthesize valuations, for which all reactivities are met.
\LongVersion{

}How to properly encode reactivities turned out rather subtle.
Let us first exemplify the complexity of the definition of reactivities.

\begin{example}\label{example:reactivity:NM}
	Consider the third reactivity in \cref{figure:reactivities} (abbreviated by NM in the following) that requires that any data transmission Meas $\rightarrow$ Navigation $\rightarrow$ Monitoring $\rightarrow$ Safeguard must always be less than 55\,ms.
	Recall that data are transmitted upon the end of a \emph{thread} period.

	We can see this reactivity as the start of a timer at the beginning of the last thread period of an execution of Navigation that completed before the end of an execution of task~T1, where T1 is such that it is the last execution of~T1 the period of which ends before the start of an execution of~Monitoring; then, the timer stops following the end of the period of an execution of~T1 immediately following the end of the period of~T3 corresponding to the end of the aforementioned execution of~Monitoring.
	At the end, the timer must be less than~55\,ms.
	
	In other words, this reactivity requires that any following sequence of actions should take less than 55\,ms:
	\styleact{startT1},
	\styleact{startNavigation} followed by \styleact{endNavigation} (without any occurrence of \styleact{startNavigation} in between)
	followed by \styleact{endT1},
	followed by \styleact{startT3} (without any occurrence of \styleact{endT1} in between),
	\styleact{startMonitoring} followed by \styleact{endMonitoring} (without any occurrence of \styleact{startMonitoring} in between),
	followed by \styleact{endT3}.
\end{example}

Encoding reactivities is arguably the most technical part of our solution, and we tried multiple solutions (either incorrect or that represented a too large overhead) before converging to this solution.
Nevertheless, the solution we chose still represents a large overhead, as we will see in \cref{section:experiments}.

In our solution, each reactivity is encoded as a sort of \emph{observer} automaton~\cite{ABBL03,Andre13ICECCS}; an observer automaton observes the system behavior without interfering with it.
That is, it can read clocks, and synchronize on synchronization actions, but without impacting the rest of the systems; in particular, it must be non-blocking (except potentially once the property verified by the observer is violated).
In addition, an observer often reduces to \emph{reachability} analysis: the property encoded by the observer is violated iff a special location of the observer is reachable.

Each reactivity automaton uses a single (local) clock used to check the reactivity constraint, and synchronizes with the rest of the system on synchronization labels encoding the start and end of processings and tasks.\todo{note pour dire qu'on les a pas tous mis ailleurs}

In fact, we deviate from the principle of observer automaton by allowing it to block in some cases.
Indeed, a key point in the definition of reactivities in our problem is the communication between threads as exemplified in \cref{example:reactivity:NM}.
In order to allow a generic solution for reactivities, and due to the fact that some timing parameters are unknown, we cannot make assumptions on the respective ordering of processings \wrt{} each other.
Therefore, when a given processing is faster than another one (\eg{} Navigation is faster than Guidance), it is not a priori possible to know which instance of the fast processing (\eg{} Navigation) will effectively transmit its data to the following slower processing (\eg{} Monitoring).
As a consequence, our observer will non-deterministically ``guess'' from which instance of the slower processing to start its timer: this is achieved by a non-deterministic choice in the initial location of the automaton.
If the guess is wrong, the observer ``blocks'' the system (impossibility to fire a transition or let time elapse).

\begin{example}\label{example:reactivity:NM2}
	Consider again reactivity NM.
	Consider a given instance of Navigation.
	If a second full instance of Navigation (including the end of thread T1) is observed before the start of~T2, our observer made a wrong guess, and the observer clock is not measuring a proper reactivity, as the instance of Navigation on which the clock should be started must be the last completed instance before the start of~T2.
	In that case, the observer simply blocks.
\end{example}

Note that, while blocking is usually not an admissible feature of observer automata, this is harmless in this case as, due to the non-deterministic guess and the fact that model checking explores all choices, all possible behaviors of the system are still explored by our solution.

\todo{préciser que les ECNA seraient un bon formalisme pour récupérer le dernier appel ?!}

\subsection{Observer construction}

Therefore, our solution consists in translating the sequence of starting and ending actions of threads and processings following the definition of the reactivities, while forbidding some actions in some locations to ensure the proper encoding of the definition of thread communication and reactivities.
In addition, a clock measuring the reactivity is started upon the (non-deterministic) activation of the first thread, and is checked against the reactivity nominal maximum time upon completion of the last thread.
If this maximum time constraint is violated, the observer enters a special ``bad'' location.
This observer violation location is added to the list of ``bad'' locations in \cref{ss:reachability} when performing reachability synthesis.

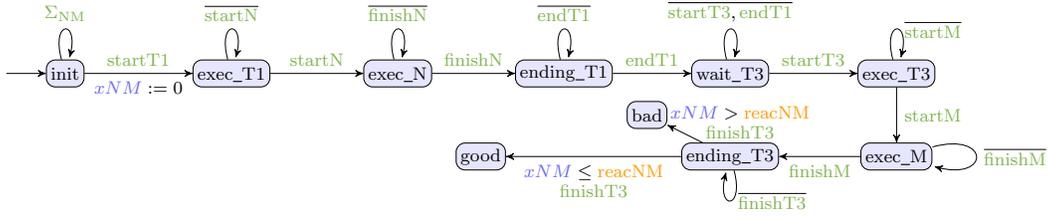
\begin{figure*}
	\centering
	\scalebox{.85}{
\begin{tikzpicture}[scale=1.3, auto, ->, >=stealth']

	\node[location, initial] at (0,0) (init) [align=center] {\styleloc{init}};

	\node[location] at (2, 0) (execT1) [align=center] {\styleloc{exec\_T1}};

	\node[location] at (4, 0) (execN) [align=center] {\styleloc{exec\_N}};

	\node[location] at (6, 0) (endingT1) [align=center] {\styleloc{ending\_T1}};

	\node[location] at (8, 0) (waitT3) [align=center] {\styleloc{wait\_T3}};

	\node[location] at (10, 0) (execT3) [align=center] {\styleloc{exec\_T3}};

	\node[location] at (10, -1) (execM) [align=center] {\styleloc{exec\_M}};

	\node[location] at (8, -1) (endingT3) [align=center] {\styleloc{ending\_T3}};


	\node[location] at (5, -1) (good) [align=center] {\styleloc{good}};

	\node[location] at (7, -.5) (bad) [align=center] {\styleloc{bad}};


	\path (init) edge node[above]{\styleact{startT1}} node[below]{$\styleclock{xNM} := 0$} (execT1);
	\path (init) edge[loop above] node[above]{\styleact{\Actions_{NM}}} (init);

	\path (execT1) edge node[auto]{\styleact{startN}} (execN);
	\path (execT1) edge[loop above] node[above]{\allActExcept{\styleact{startN}}} (execT1);
	
	\path (execN) edge node[auto]{\styleact{finishN}} (endingT1);
	\path (execN) edge[loop above] node[above]{\allActExcept{\styleact{finishN}}} (execN);
	
	\path (endingT1) edge node[auto]{\styleact{endT1}} (waitT3);
	\path (endingT1) edge[loop above] node[above]{\allActExcept{\styleact{endT1}}} (endingT1);
	
	\path (waitT3) edge node[auto]{\styleact{startT3}} (execT3);
	\path (waitT3) edge[loop above] node[above]{\allActExcept{\styleact{startT3}, \styleact{endT1}}} (waitT3);
	
	\path (execT3) edge node[auto]{\styleact{startM}} (execM);
	\path (execT3) edge[loop above] node[right]{\allActExcept{\styleact{startM}}} (execT3);
	
	\path (execM) edge node[auto]{\styleact{finishM}} (endingT3);
	\path (execM) edge[loop right] node[right]{\allActExcept{\styleact{finishM}}} (execM);
	
	\path (endingT3) edge node[auto,align=center]{$\styleclock{xNM} \leq \styleparam{reacNM}$\\\styleact{finishT3}} (good);
	\path (endingT3) edge[loop below] node[right]{\allActExcept{\styleact{finishT3}}} (endingT3);
	
	
	\path (endingT3) edge node[right,yshift=.5em, xshift=-1em,align=center]{$\styleclock{xNM} > \styleparam{reacNM}$\\\styleact{finishT3}} (bad);
\end{tikzpicture}
}
	
	\caption{Encoding reactivity Navigation $\rightarrow$ Monitoring}
	\label{figure:reactivityNM}
\end{figure*}

%

\begin{example}\label{example:reactivity:NM-PTA}
	We give the observer automaton corresponding to reactivity NM in \cref{figure:reactivityNM}.
	We abbreviate\LongVersion{ in \cref{figure:reactivityNM}} the names of processings (N and~M stand for Navigation and Monitoring respectively).
	The only clock is \styleclock{xNM} while \styleparam{reacNM} denotes the maximum nominal reactivity for NM (55\,ms in our setting).
	$\Actions_{NM}$ stands for this automaton alphabet; given $\action \in \Actions_{NM}$, $\allActExcept{\action}$ denotes $\Actions_{NM} \setminus \{ \action \}$ (we extend this notation to sets of actions).
	In addition, whenever $\styleclock{xNM} > \styleparam{reacNM}$ occurs in any location (except the initial location), a transition leads to \styleloc{bad} (not depicted in \cref{figure:reactivityNM} for sake of clarity).
	
		
	The non-deterministic choice is encoded in the initial location where, upon action~\styleact{startT1}, the automaton can either self-loop in \styleloc{init}, or go to \styleloc{startT1} to try to measure the reactivity from this instance of~T1.
	The blocking is encoded by the absence of transition labeled with \styleact{endT1} in location \styleloc{wait\_T3} (an alternative is to synchronize on \styleact{endT1} to a sink location that also blocks time elapsing).
\end{example}

Both remaining reactivities in \cref{figure:reactivities} follow easily from this scheme: the first reactivity (Navigation $\rightarrow$ Guidance $\rightarrow$ Control) follows the same principle for Navigation and Guidance, and is immediately followed by a third check for Control, while the second reactivity (Navigation $\rightarrow$ Control) is simpler as both Navigation and Control are on the same thread.

\subsection{Compositional verification and synthesis}

Due to the non-deterministic choice, the verification of the reactivities entails a clear overhead to the verification (see \cref{section:experiments}).
Verifying all three reactivities can be naturally done by adding the three observer automata to the same system, and performing synthesis on the composition of all these automata.

However, we claim that this can be done in a \emph{compositional} fashion.
Indeed, checking reactivities is checking that a constraint is met for all executions; this can be seen as a global invariant.
Therefore, checking that these three invariants are valid can be done separately.
In the non-parametric case, we will perform three different verifications of the system, with only one reactivity automaton at a time.
Then, if the ``bad'' locations are unreachable for the three different verifications, then the system is schedulable and the reactivities are met.
In the case of synthesis, we will \emph{intersect} the result of the synthesis applied to the three parametric models.

This compositional analysis comes in contrast with many works on scheduling, where compositionality is hard to achieve.

\section{Experiments}\label{section:experiments}
    
\subsection{Experimental environment}
We modeled our network of PSA in the \imitator{} input language~\cite{AFKS12}.
\imitator{} is a parametric model checker taking as input networks of PSA extended with useful features such as synchronization actions and discrete variables.
Synthesis can be performed using various properties including reachability---which is the feature we use here.
When \imitator{} terminates (which is not guaranteed in theory), the tool is often able to infer whether the result is exact (sound and complete);
all analyses mentioned in this manuscript terminate with an exact result.


The translation effort was manual due to the specificity of our solution (with the exception of the scheduler, for which we started from an automated generator).
However, we tried to keep our translation as systematic as possible to allow for a future automated generation from the problem input data.
We made intensive use of clock resets and stopwatches for clocks not necessary at some points, in order to let \imitator{} apply inactive clock reductions.

All experiments were conducted using \imitator{} 2.10.4 ``Butter Jellyfish'' on an ASUS X411UN Intel Core$\texttrademark$ i7-8550U 1.80\,GHz 
	with 8\,GiB memory running Linux Mint~19 64\,bits.\footnote{%
	Sources, \LongVersion{binaries, }models and results are available at \url{imitator.fr/static/ACSD19/}
}

\subsection{Verification and synthesis without reactivities}

In order to evaluate the overhead of the satisfaction of the reactivities, we first run analyses \emph{without} reactivities.

\paragraph{Non-parametric model}
First, a non-parametric analysis shows that the bad locations are unreachable, and therefore the system is schedulable under the nominal values given in \cref{figure:example_control_system,figure:example_wcet}.
We give in \LongVersion{\cref{fig:GNC-with-Cheddar} in \cref{appendix:cheddar}}\ShortVersion{\cite{ACFJL19report}} the Gantt chart (obtained with \cheddar{}~\cite{Cheddar}) of this entirely instantiated model.
Computation times of all the analyses without reactivities are given in \cref{table:time-no-reac}.

\begin{table}
	\centering
	\caption{Computation times without reactivities}
	\begin{tabular}{| l | r | }
		\hline
		\cellHeader{Analysis} & \cellHeader{Time (s)} \\
		\hline
		No parameter & 3.086\\
		\hline
		Parametric offsets & 95.807\\
		\hline
		Parametric deadlines & 17.689\\
		\hline
	\end{tabular}
	\label{table:time-no-reac}
\end{table}

\paragraph{Parameterized offsets}
We then parameterize offsets, \ie{} we seek admissible offsets for which the system is schedulable.
The constraint is given in \LongVersion{\cref{appendix:parametric-offsets}}\ShortVersion{\cite{ACFJL19report}}.
\LongVersion{We can see that, while several conditions for schedulability are given, at least one offset must be 0 to ensure schedulability.}

\paragraph{Parameterized deadlines}
We then parameterize deadlines, \ie{} we seek admissible deadlines for which the system is schedulable.
The constraint is:
 $ \deadlineTtwo{} \in [11, 20]
 \land\ \deadlineTone{} \in [4, 5]
 \land\ \deadlineTthree{} = 60$.
That is, the deadline of~T3 is strict, while T1 and~T2 can be relaxed while preserving schedulability.


\subsection{Compositional verification of reactivities}

We then solve the scheduling verification and scheduling synthesis problems with reactivities, using two methods:
\begin{itemize}
	\item monolithic verification: all three reactivity automata are included in the model;
	\item compositional verification: we verify sequentially three different models, each of them including all automata modeling the system, but only one reactivity at a time.
\end{itemize}

\begin{table}
	\centering
	\caption{Computation times with reactivities}
	\begin{tabular}{| l | r | r | }
		\hline
		\cellHeader{Analysis} & \cellHeader{Monolithic (s)}  & \cellHeader{Compositional (s)} \\
		\hline
		No parameter & 109.404 & 40.092\\
		\hline
		Parametric offsets & 2303.975 & 2278.363\\
		\hline
		Parametric deadlines & 637.169 & 331.272\\
		\hline
	\end{tabular}
	\label{table:time-reac}
\end{table}

We give the various computation times in \cref{table:time-reac}.
A full version of this table, including the overhead incurred by each reactivity, is given in \LongVersion{\cref{table:time-reac-full} in \cref{appendix:time-reac-full}}\ShortVersion{\cite{ACFJL19report}}.
\cref{table:time-reac} shows the interest of the compositional verification over monolithic verification, as the computation time is divided by a factor~2.

\section{Comparison with other tools}\label{section:comparison}

We perform a comparison with two other well-known tools, one from the real-time system community, namely \cheddar{}~\cite{Cheddar}, and one from the timed automata community, namely \uppaal{}~\cite{LPY97}.
Both tools cannot handle parameters nor consider partially specified problems, and therefore can only solve the scheduling verification problem.
Therefore, in this section, we consider the instantiated version of the system according to the nominal values given in \cref{figure:example_control_system,figure:example_wcet}.
In addition, to the best of our knowledge, \cheddar{} cannot test the reactivities.

\paragraph{Comparison with \cheddar{}}

\cheddar{} is a real-time scheduling tool distributed under the GPL license.
\cheddar{} is used to model software architectures of real-time systems and to check if the system is schedulable.

\LongVersion{We checked the system's schedulability using \cheddar{} when the system is instantiated (\ie{} all offsets are initialized to 0 and the deadline of each thread equal to the period).
We have indicated the period, the execution time and deadline of each processings.}

As result, \cheddar{} proves that the system is schedulable and there are no deadline missed in the computed scheduling.
\LongVersion{%
	In this solution, the number of context switches per period of~T3 is~29 and the number of preemptions is~8.\todo{Jawher: je ne comprends pas:  29 switches et 8 préemptions, OK, mais en combien de temps…? C'est-à-dire quand est-ce que Cheddar s'arrête de calculer ? (le système est infini, donc Cheddar doit s'arrêter quelque part à un endroit où il trouve que c'est suffisant d'arrêter là sans explorer le reste)}%
		\jj{Ce modèle rassemble un peu le modèle instancié de GNC, la différence est que: dans ce modèle Control s'exécute pendant le cycle pair (cycle impair dans GNC). Les 29 switches et 8 préemptions (3 préemptions de Monitoring et 5 préemptions de Guidance ) sont calculés pendant 60 ms(periode de Guidance). J'ai ajouté un autre modèle avec Cheddar où control s'exécute pendanr le cycle impair comme le modèle GNC et qui donne aussi 29 switches et 8 préemptions}%
		\ea{Merci. Mais c'est toi qui a mis la limite de 60, ou c'est \cheddar{}?}
	\jj{Oui, c'est Cheddar qui a choisit la limite de 60}
}

\LongVersion{%
\cheddar{} cannot give a solution to the scheduling synthesis problem since it only works with instantiated systems, so we cannot determine offsets and deadlines, and also it does not deal with reactivities.
}

\paragraph{Comparison with \uppaal{}}

We also compare the obtained results using \imitator{} with \uppaal{} results.
We wrote a \uppaal{} model identical to the \imitator{} model---with instantiated parameters.
\LongVersion{%

}%
As result, \uppaal{} proves that the instantiated system is schedulable, both without and with reactivities.

\paragraph{Summary of comparisons}

We give the computation times without reactivities in \cref{table:comparison-no-parameters}.
Clearly, without parameters, \cheddar{} or \uppaal{} should be used.
However none of these tools cope with uncertain constants.


\begin{table}
	\centering
	\caption{Computation times without parameters}
	\begin{tabular}{| l | r | r | }
		\hline
		\cellHeader{Analysis} & \cellHeader{Without reactivities (s)} & \cellHeader{With reactivities (s)} \\
		\hline
		\cheddar{} & $<0.1$ & \cellNA{}\\
		\hline
		\imitator{} & 3.086 & 109.404 \\
		\hline
		\uppaal{} & 0.002 & 0.003\\
		\hline
	\end{tabular}
	\label{table:comparison-no-parameters}
\end{table}

%
%

\section{Conclusion and perspectives}\label{section:conclusion}


We proposed an approach to synthesize timing valuations ensuring schedulability of the flight control of a space launcher.
A key issue is to ensure that the system reactivities are met---for which we proposed a compositional solution.

\paragraph*{Future works}
We omitted an element of the problem: the switch between two threads has a CPU cost due to the copy of data between the contexts of each thread which is in our example $500\,\mu{}s$.
This can be added by modifying the scheduler automaton; a more interesting outcome will be to \emph{synthesize} the maximum admissible switch time (possibly depending on other parameters) that still ensures schedulability.

Due to the efficiency gap of an order of magnitude, combining some non-parametric analyses (\eg{} with \uppaal{} or \cheddar{}) with parametric analyses (\imitator{}) would be an interesting future work.

The harmonic periods are a strong assumption of the problem.
Tuning our solution to benefit from this assumption is on our agenda.

We envisage two tracks for our longer-term future works:
\begin{itemize}
	\item Generalizing the flight control scheduling problem by automatically synthesizing the allocations of processings on threads.
		This generalization raises first the issue of modeling such problematic (how to model an allocation with a parameter) and second the classical combinatorial explosion of states.
	\item Applying this approach to the automatic synthesis of the launcher sequential, \ie{} of the scheduling of all the \LongVersion{system }events necessary to fulfill a mission: ignition and shut-down of stages, release of firing, release of payloads, etc.
\end{itemize}



\ifdefined\VersionLong
	\bibliographystyle{alpha} 
	\newcommand{\CCIS}{Communications in Computer and Information Science}
	\newcommand{\ENTCS}{Electronic Notes in Theoretical Computer Science}
	\newcommand{\FMSD}{Formal Methods in System Design}
	\newcommand{\IJFCS}{International Journal of Foundations of Computer Science}
	\newcommand{\IJSSE}{International Journal of Secure Software Engineering}
	\newcommand{\JLAP}{Journal of Logic and Algebraic Programming}
	\newcommand{\JLC}{Journal of Logic and Computation}
	\newcommand{\LMCS}{Logical Methods in Computer Science}
	\newcommand{\LNCS}{Lecture Notes in Computer Science}
	\newcommand{\RESS}{Reliability Engineering \& System Safety}
	\newcommand{\STTT}{International Journal on Software Tools for Technology Transfer}
	\newcommand{\TCS}{Theoretical Computer Science}
	\newcommand{\ToPNoC}{Transactions on Petri Nets and Other Models of Concurrency}
	\newcommand{\TSE}{IEEE Transactions on Software Engineering}
\else
	\bibliographystyle{IEEEtran} 
	\newcommand{\CCIS}{CCIS}
	\newcommand{\ENTCS}{ENTCS}
	\newcommand{\FMSD}{FMSD}
	\newcommand{\IJFCS}{IJFCS}
	\newcommand{\IJSSE}{IJSSE}
	\newcommand{\JLAP}{JLAP}
	\newcommand{\JLC}{JLC}
	\newcommand{\LMCS}{LMCS}
	\newcommand{\LNCS}{LNCS}
	\newcommand{\RESS}{RESS}
	\newcommand{\STTT}{STTT}
	\newcommand{\TCS}{TCS}
	\newcommand{\ToPNoC}{ToPNoC}
	\newcommand{\TSE}{TSE}
\fi
\bibliography{Ariane}

\LongVersion{

\newpage
\appendix

\subsection{Parametric analyses without reactivities}

\subsubsection{Parametric offsets}\label{appendix:parametric-offsets}

See \cref{table:result-parametric-offsets}.


\begin{figure}[H]
	\centering
	\begin{tabular}{| l |}
	\hline
$ 5 >= \offsetTtwo{}$\\
$\land\ \offsetTthree{} + 5 > \offsetTtwo{}$\\
$\land\ \offsetTthree{} >= 0$\\
$\land\ \offsetTtwo{} >= 0$\\
$\land\ 1 >= \offsetTthree{}$\\
$\land\ \offsetTone{} = 0$\\
\textbf{OR}\\
$  \offsetTone{} >= 0$\\
$\land\ 11 >= \offsetTthree{}$\\
$\land\ \offsetTthree{} > 1 + \offsetTone{}$\\
$\land\ 4 >= \offsetTone{}$\\
$\land\ \offsetTtwo{} = 0$\\
\textbf{OR}\\
$  \offsetTthree{} > 1$\\
$\land\ 11 >= \offsetTthree{}$\\
$\land\ \offsetTtwo{} > 0$\\
$\land\ 1 >= \offsetTtwo{}$\\
$\land\ \offsetTone{} = 0$\\
\textbf{OR}\\
$  \offsetTone{} > 0$\\
$\land\ \offsetTtwo{} >= 0$\\
$\land\ 11 >= \offsetTtwo{}$\\
$\land\ 4 >= \offsetTone{}$\\
$\land\ \offsetTthree{} = 0$\\
\textbf{OR}\\
$  11 >= \offsetTtwo{}$\\
$\land\ \offsetTthree{} >= 0$\\
$\land\ \offsetTtwo{} > 9$\\
$\land\ 1 >= \offsetTthree{}$\\
$\land\ \offsetTone{} = 0$\\
\textbf{OR}\\
$  \offsetTone{} + 1 >= \offsetTthree{}$\\
$\land\ \offsetTone{} > 0$\\
$\land\ \offsetTthree{} > 0$\\
$\land\ 4 >= \offsetTone{}$\\
$\land\ \offsetTtwo{} = 0$\\
\textbf{OR}\\
$  \offsetTtwo{} > 5$\\
$\land\ 9 >= \offsetTtwo{}$\\
$\land\ \offsetTthree{} > 0$\\
$\land\ 1 >= \offsetTthree{}$\\
$\land\ \offsetTone{} = 0$\\
\textbf{OR}\\
$  \offsetTtwo{} >= 5$\\
$\land\ 9 >= \offsetTtwo{}$\\
$\land\ \offsetTone{} = 0$\\
$\land\ \offsetTthree{} = 0$\\
		\hline
	\end{tabular}
	\caption{Parametric offsets}
	\label{table:result-parametric-offsets}
\end{figure}

%

\LongVersion{
\subsubsection{Parametric offsets and deadlines}\label{appendix:parametric-offsets-deadlines}


\begin{figure}[H]
	\centering
	\begin{tabular}{| l |}
	\hline
$\deadlineTtwo{} > 11$\\
$\land\ 11 >= \offsetTthree{}$\\
$\land\ \deadlineTone{} >= 4$\\
$\land\ \offsetTthree{} > \offsetTtwo{}$\\
$\land\ 20 >= \deadlineTtwo{}$\\
$\land\ \offsetTtwo{} > 0$\\
$\land\ 5 >= \deadlineTone{}$\\
$\land\ 1 >= \offsetTtwo{}$\\
$\land\ \offsetTone{} = 0$\\
$\land\ \deadlineTthree{} = 60$\\
\textbf{OR}\\
  $\offsetTone{} > 0$\\
$\land\ \deadlineTone{} >= 4$\\
$\land\ 20 >= \deadlineTtwo{}$\\
$\land\ \offsetTthree{} > 5$\\
$\land\ \deadlineTtwo{} >= 15$\\
$\land\ 11 >= \offsetTthree{}$\\
$\land\ 5 >= \deadlineTone{}$\\
$\land\ \deadlineTone{} >= 1 +\offsetTone{}$\\
$\land\ \offsetTtwo{} = 0$\\
$\land\ \deadlineTthree{} = 60$\\
\textbf{OR}\\
  $\offsetTone{} > 0$\\
$\land\ \offsetTone{} + 1 >= \offsetTthree{}$\\
$\land\ \deadlineTtwo{} >= 15$\\
$\land\ 20 >= \deadlineTtwo{}$\\
$\land\ 4 >=\offsetTone{}$\\
$\land\ \offsetTthree{} >= 0$\\
$\land\ \deadlineTone{} = 5$\\
$\land\ \offsetTtwo{} = 0$\\
$\land\ \deadlineTthree{} = 60$\\
\textbf{OR}\\
   $\deadlineTtwo{} > 11$\\
$\land\ 11 >= \offsetTthree{}$\\
$\land\ \deadlineTone{} >= 4$\\
$\land\ \offsetTthree{} > \deadlineTone{}$\\
$\land\ 20 >= \deadlineTtwo{}$\\
$\land\ 5 >= \deadlineTone{}$\\
$\land\ \offsetTtwo{} = 0$\\
$\land\ \offsetTone{} = 0$\\
$\land\ \deadlineTthree{} = 60$\\
\textbf{OR}\\
  $ \deadlineTtwo{} > 11$\\
$\land\ 1 >= \offsetTthree{}$\\
$\land\ \deadlineTtwo{} >= 10 + \offsetTtwo{}$\\
$\land\ \offsetTtwo{} >= 1$\\
$\land\ 20 >= \deadlineTtwo{}$\\
$\land\ \deadlineTone{} >= \offsetTtwo{}$\\
$\land\ 5 > \deadlineTone{}$\\
$\land\ \offsetTthree{} >= 0$\\
$\land\ \deadlineTone{} >= 4$\\
$\land\ \offsetTone{} = 0$\\
$\land\ \deadlineTthree{} = 60$\\
\textbf{OR}\\
   $20 >= \deadlineTtwo{}$\\
$\land\ \deadlineTone{} >= 4$\\
$\land\ \offsetTone{} > 0$\\
$\land\ \deadlineTone{} >= 1 +\offsetTone{}$\\
$\land\ \deadlineTtwo{} > 11$\\
$\land\ \offsetTthree{} > 1 +\offsetTone{}$\\
$\land\ 5 >= \deadlineTone{}$\\
$\land\ 5 >= \offsetTthree{}$\\
$\land\ \offsetTtwo{} = 0$\\
$\land\ \deadlineTthree{} = 60$\\
\textbf{OR}\\
   $offsetT1 >= 0$\\
$\land\ \deadlineTtwo{} > 11$\\
$\land\ \offsetTone{} + 1 >= \offsetTthree{}$\\
$\land\ \deadlineTone{} >= 4$\\
$\land\ \deadlineTtwo{} >= 11 +\offsetTone{}$\\
$\land\ 20 >= \deadlineTtwo{}$\\
$\land\ \offsetTthree{} >= 0$\\
$\land\ 5 > \deadlineTone{}$\\
$\land\ \deadlineTone{} >= 1 +\offsetTone{}$\\
$\land\ \offsetTtwo{} = 0$\\
$\land\ \deadlineTthree{} = 60$\\

\hline
\end{tabular}
\end{figure}
\begin{figure}[H]
\centering
	\begin{tabular}{| l |}
	\hline
\textbf{OR}\\
   $\offsetTtwo{} > 5$\\
$\land\ 20 >= \deadlineTtwo{}$\\
$\land\ \offsetTone{} >= 0$\\
$\land\ \deadlineTone{} >= 4$\\
$\land\ \deadlineTtwo{} >= 10 + \offsetTtwo{}$\\
$\land\ 5 >= \deadlineTone{}$\\
$\land\ \deadlineTone{} >= 1 +\offsetTone{}$\\
$\land\ \deadlineTtwo{} >= 19$\\
$\land\ \offsetTthree{} = 0$\\
$\land\ \deadlineTthree{} = 60$\\
\textbf{OR}\\
  $ \deadlineTtwo{} >= 11 +\offsetTone{}$\\
$\land\ \deadlineTtwo{} > 11$\\
$\land\ \offsetTtwo{} > 0$\\
$\land\ \offsetTone{} >= 0$\\
$\land\ 5 >= \deadlineTone{}$\\
$\land\ 20 >= \deadlineTtwo{}$\\
$\land\ 4 >\offsetTone{}$\\
$\land\ \offsetTtwo{} + 5 >\offsetTone{} + \deadlineTone{}$\\
$\land\ \deadlineTone{} >= 1 +\offsetTone{}$\\
$\land\ \offsetTone{} + 1 > \offsetTtwo{}$\\
$\land\ \offsetTtwo{} >=\offsetTone{}$\\
$\land\ \deadlineTone{} >= 4$\\
$\land\ \offsetTthree{} = 0$\\
$\land\ \deadlineTthree{} = 60$\\
\textbf{OR}\\
  $ \deadlineTtwo{} > 11$\\
$\land\ 19 > \deadlineTtwo{}$\\
$\land\ \offsetTtwo{} >= 1$\\
$\land\ 5 > \offsetTtwo{}$\\
$\land\ \offsetTthree{} > 0$\\
$\land\ \deadlineTtwo{} >= 10 + \offsetTtwo{}$\\
$\land\ 1 >= \offsetTthree{}$\\
$\land\ \offsetTone{} = 0$\\
$\land\ \deadlineTone{} = 5$\\
$\land\ \deadlineTthree{} = 60$\\
\textbf{OR}\\
  $ \deadlineTtwo{} > 11$\\
$\land\ \deadlineTone{} >= 4$\\
$\land\ 5 >= \deadlineTone{}$\\
$\land\ \offsetTthree{} > 0$\\
$\land\ 1 > \offsetTtwo{}$\\
$\land\ 20 >= \deadlineTtwo{}$\\
$\land\ \offsetTtwo{} >= \offsetTthree{}$\\
$\land\ \offsetTone{} = 0$\\
$\land\ \deadlineTthree{} = 60$\\
\textbf{OR}\\
   $19 > \deadlineTtwo{}$\\
$\land\ \offsetTthree{} >= 0$\\
$\land\ \offsetTtwo{} > \deadlineTone{}$\\
$\land\ \deadlineTtwo{} >= 10 + \offsetTtwo{}$\\
$\land\ 5 > \offsetTtwo{}$\\
$\land\ 1 >= \offsetTthree{}$\\
$\land\ \deadlineTone{} >= 4$\\
$\land\ \offsetTone{} = 0$\\
$\land\ \deadlineTthree{} = 60$\\
		\hline
	\end{tabular}
	\label{table:result-parametric-offsets+deadlines}
\end{figure}

}

\subsection{Parametric analyses with reactivities}

\subsubsection{Precise computation times}\label{appendix:time-reac-full}

The computation times with reactivities giving the details of the compositional analysis is given in \cref{table:time-reac-full}.

\begin{table*}
	\centering
	\caption{Computation times with reactivities (full)}
	\begin{tabular}{| l | r | r | r | r | r | }
		\hline
		\cellHeader{Analysis} & \cellHeader{Monolithic (s)}  & \cellHeader{NGC (s)}  & \cellHeader{NC (s)}  & \cellHeader{NM (s)} & \cellHeader{Compositional (s)} \\
		\hline
		No parameter & 109.404 & 21.427 & 3.444 & 15.221 & 40.092 \\
		\hline
		Parametric offsets & 2303.975 & 1111.916 & 210.784 & 955.663  & 2278.363\\
		\hline
		Parametric deadlines & 637.169 & 172.962 & 28.540 & 129.770 & 331.272\\
		\hline
	\end{tabular}
	\label{table:time-reac-full}
\end{table*}

\subsection{Parametric analyses with switch time without reactivities}

\subsubsection{Parametric offsets}\label{appendix:parametric-offsets-with-switch-time}
See \cref{table:result-parametric-offsets-with-switch-time}.

\begin{figure}[H]
	\centering
	\begin{tabular}{| l |}
	\hline
$ \offsetTthree{} >= 10$\\
$\land\ 7 >= 2*\offsetTone{}$\\
$\land\ 2*\offsetTone{} > 5 + 2*\offsetTtwo{}$\\
$\land\ 23 >= 2*\offsetTthree{}$\\
$\land\ \offsetTone{} >= 3$\\
$\land\ 2*\offsetTtwo{} + 7 > 2*\offsetTone{}$\\
$\land\ \offsetTtwo{} >= 0$\\
$\land\ \offsetTone{} > 2 + \offsetTtwo{}$\\
\textbf{OR}\\
$  \offsetTtwo{} + 2*\offsetTthree{} + 3 > 3*\offsetTone{}$\\
$\land\ \offsetTtwo{} >= 0$\\
$\land\ 2*\offsetTthree{} > 1 + 2*\offsetTone{}$\\
$\land\ 19 >= 2*\offsetTthree{}$\\
$\land\ 2*\offsetTtwo{} + 7 > 2*\offsetTone{}$\\
$\land\ \offsetTone{} >= 3 + \offsetTtwo{}$\\
$\land\ 7 >= 2*\offsetTone{}$\\
\textbf{OR}\\
$  23 >= 2*\offsetTtwo{}$\\
$\land\ \offsetTthree{} >= 0$\\
$\land\ \offsetTtwo{} >= 5$\\
$\land\ 1 >= \offsetTthree{}$\\
$\land\ \offsetTone{} = 0$\\
\textbf{OR}\\
$  19 >= 2*\offsetTthree{}$\\
$\land\ \offsetTone{} >= 0$\\
$\land\ \offsetTthree{} > 5$\\
$\land\ \offsetTtwo{} > 1 + \offsetTone{}$\\
$\land\ 2 > \offsetTtwo{}$\\
\textbf{OR}\\
$  \offsetTone{} >= 0$\\
$\land\ \offsetTtwo{} > 0$\\
$\land\ \offsetTthree{} > 5 + \offsetTtwo{}$\\
$\land\ \offsetTtwo{} >= \offsetTone{}$\\
$\land\ \offsetTone{} + 1 >= \offsetTtwo{}$\\
$\land\ 1 > 2*\offsetTone{}$\\
$\land\ 19 >= 2*\offsetTthree{}$\\
\textbf{OR}\\
$  \offsetTtwo{} >= 5$\\
$\land\ 23 >= 2*\offsetTtwo{}$\\
$\land\ \offsetTone{} >= 0$\\
$\land\ \offsetTthree{} > 1 + \offsetTone{}$\\
$\land\ 3 >= 2*\offsetTthree{}$\\
\textbf{OR}\\
$  \offsetTthree{} > 3$\\
$\land\ \offsetTthree{} >= 3 + \offsetTtwo{}$\\
$\land\ 2*\offsetTthree{} > 5 + 2*\offsetTtwo{}$\\
$\land\ \offsetTone{} + \offsetTthree{} > 6 + 2*\offsetTtwo{}$\\
$\land\ \offsetTone{} >= 3 + \offsetTtwo{}$\\
$\land\ \offsetTtwo{} >= 0$\\
$\land\ 2*\offsetTone{} + 1 >= 2*\offsetTthree{}$\\
$\land\ 7 >= 2*\offsetTone{}$\\
\textbf{OR}\\
$  2*\offsetTtwo{} > 7$\\
$\land\ 3*\offsetTtwo{} > 4$\\
$\land\ \offsetTtwo{} + 8 > 0$\\
$\land\ 10 > \offsetTtwo{}$\\
$\land\ 2*\offsetTone{} = 7$\\
$\land\ \offsetTthree{} = 0$\\
		\hline
	\end{tabular}
	\caption{Parametric offsets for model with switch time}
	\label{table:result-parametric-offsets-with-switch-time}
\end{figure}

\subsubsection{Parametric deadlines}\label{appendix:parametric-offsets-deadlines-with-switch-time}
See \cref{table:result-parametric-deadlines-with-switch-time}.
\begin{figure}[H]
	\centering
	\begin{tabular}{| l |}
	\hline
	
$2*\deadlineTtwo{} >= 9$\\
$\land\ 2*\deadlineTone{} >= 9$\\
$\land\ 5 >= \deadlineTone{}$\\
$\land\ 20 >= \deadlineTtwo{}$\\
$\land\ \deadlineTthree{} = 60$\\
		\hline
	\end{tabular}
	\caption{Parametric deadlines for model with switch time}
	\label{table:result-parametric-deadlines-with-switch-time}
\end{figure}

\subsection{Scheduling with \cheddar{}}\label{appendix:cheddar}

We give in \cref{fig:GNC-with-Cheddar} 
	(obtained with \cheddar{}~\cite{Cheddar}) the Gantt chart of this entirely instantiated model.

\begin{figure*}[htbp!]
	\centering
  \includegraphics[width=.8\textwidth]{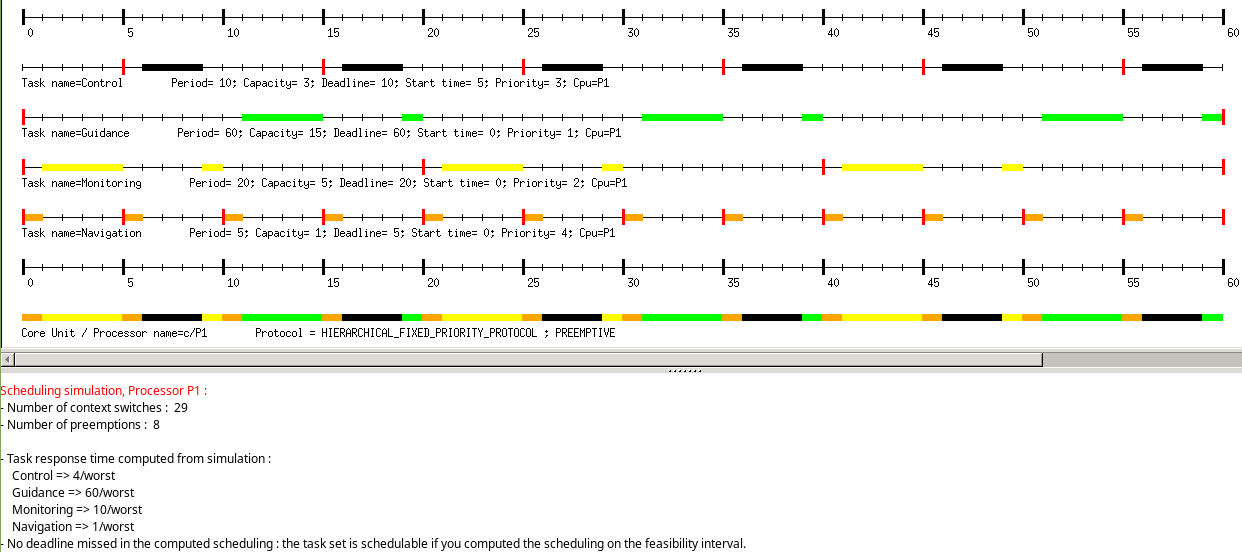}
  \caption{scheduling GNC without reactivities using \cheddar{}}
  \label{fig:GNC-with-Cheddar}
\end{figure*}

\LongVersion{%
In \cref{fig:Cheddar}, we present an example of a scheduling analysis of a system with \cheddar{}.\ea{là, il vaudrait mieux mettre l'ordonnancement de notre système (instancié, donc), plutôt que ``un'' système arbitraire}
\todo{Jawher: seulement si tu as le temps, peux-tu refaire \texttt{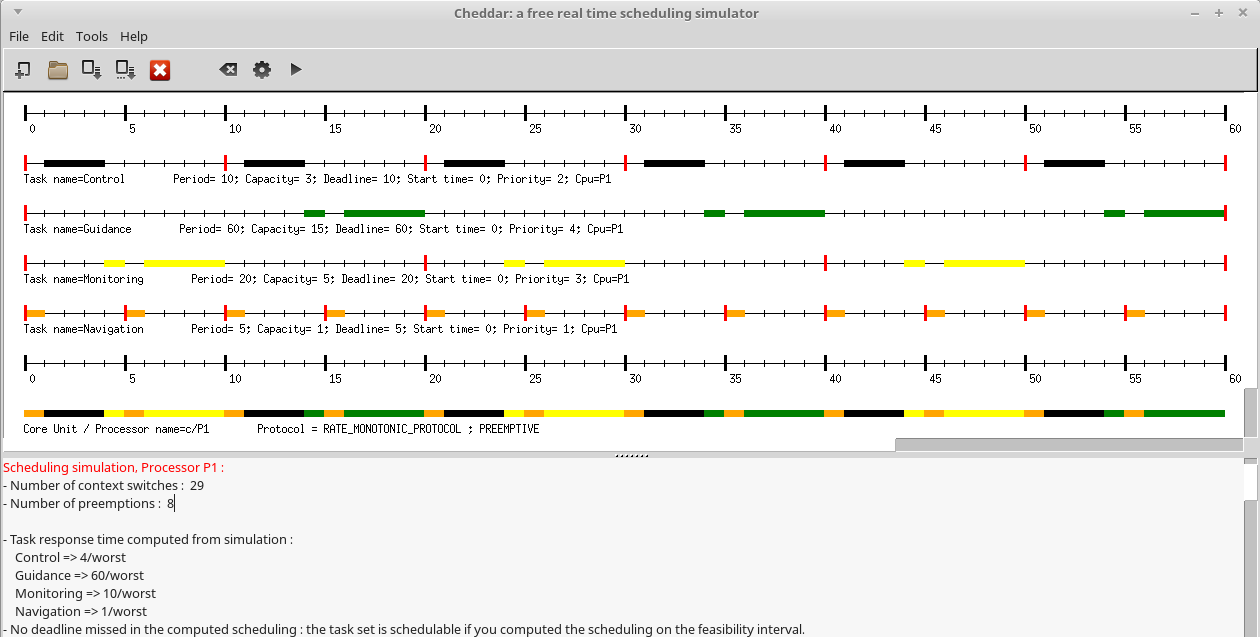} avec une capture de notre modèle à la place de ce modèle ?}
\jj{J'ai ajouté la \cref{fig:GNC-with-Cheddar} qui présente le modèle GNC avec \cheddar{}}
\begin{figure*}[htbp!]
	\centering
  \includegraphics[width=.8\textwidth]{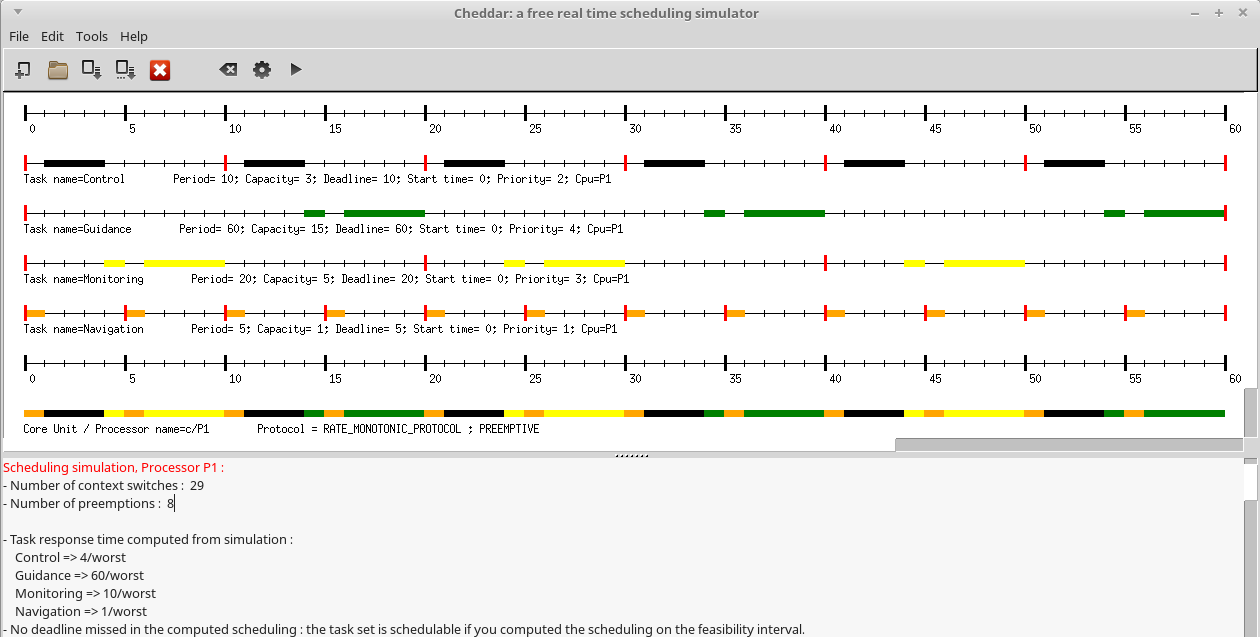}
  \caption{scheduling tasks with \cheddar{}}
  \label{fig:Cheddar}
\end{figure*}

}

%
%

}

\ea{là j'ai supprimé pas mal de choses qui ne compilaient pas, à remettre}

\end{document}